\documentclass[a4paper,fleqn]{cas-dc}

\usepackage[authoryear,longnamesfirst]{natbib}

\def\tsc#1{\csdef{#1}{\textsc{\lowercase{#1}}\xspace}}
\tsc{WGM}
\tsc{QE}
\tsc{EP}
\tsc{PMS}
\tsc{BEC}
\tsc{DE}

\def\bq{\begin{equation}}
\def\eq{\end{equation}}
\def\bqy{\begin{eqnarray}}
\def\eqy{\end{eqnarray}}
\usepackage{amssymb}
\usepackage{amsmath}
\usepackage{latexsym}

\usepackage[switch]{lineno}

\usepackage{url}
\usepackage{xcolor}
\definecolor{newcolor}{rgb}{.8,.349,.1}

\usepackage{hyperref}
\usepackage{rotating}

\usepackage{caption}
\usepackage{subcaption}
\ExplSyntaxOn \cs_gset:Npn \__first_footerline: { \group_begin: \small \sffamily \__short_authors: \group_end: } \ExplSyntaxOff

\begin{document}
\let\WriteBookmarks\relax
\def\floatpagepagefraction{1}
\def\textpagefraction{.001}

\shorttitle{Swarming Proxima Centauri}

\shortauthors{T.M. Eubanks et~al.}

\title [mode = title]{Swarming Proxima Centauri:
Optical Communications Over Interstellar Distances}           

%
\author[1]{T. Marshall Eubanks}[type=editor,
                    auid=000,bioid=1,orcid=0000-0001-9543-0414]

\cormark[1]

\ead{tme@space-initiatives.com}



\affiliation[1]{organization={Space Initiatives Inc},
    addressline={106 Mahood Avenue}, 
    city={Princeton},
    postcode={24740}, 
    state={West Virginia},
    country={USA}}

\author[1]{W. Paul Blase}

\author[2]{Andreas M. Hein}[orcid=0000-0003-1763-6892]
\affiliation[2]{organization={Luxembourg University},
    city={Luxembourg},
    country={Luxembourg}}

\author[3]{Adam Hibberd}[orcid=0000-0003-1116-576X]
\affiliation[3]{organization={Initiative for Interstellar Studies (i4is)},
    addressline={ 27/29 South Lambeth Road}, 
    city={London},
    postcode={SW8 1SZ},
    country={United Kingdom}}

\author[4]{Robert G. Kennedy III}[orcid=0000-0003-3924-7935]
\affiliation[4]{organization={Institute for Interstellar Studies (i4is US)},
    addressline={112 Mason Lane}, 
    city={Oak Ridge},
     state={Tennessee},    
    postcode={37830},
    country={USA}}

\cortext[cor1]{Corresponding author}

\begin{abstract}
Interstellar communications are achievable with gram-scale spacecraft using swarm techniques introduced herein if an adequate energy source, clocks and a suitable communications protocol exist.  The essence of our approach to the Breakthrough Starshot challenge is to launch a long string of 100s of gram-scale interstellar probes at 0.2c in a firing campaign up to a year long, maintain continuous contact with them (directly amongst each other and via Earth utilizing the launch laser), and gradually, during the 20-year cruise, dynamically coalesce the long string into a lens-shaped mesh network $\sim$100,000 km across centered on the target planet Proxima b at the time of fly-by. \\

In-flight formation would be accomplished using the ``time on target'' technique of grossly modulating the initial launch velocity between the head and the tail of the string, and combined with continual fine control or ``velocity on target'' by adjusting the attitude of selected probes, exploiting the drag imparted by the ISM. \\

Such a swarm could tolerate significant attrition, e.g., by collisions enroute with interstellar dust grains, thus mitigating the risk that comes with ``putting all your eggs in one basket''.  It would also enable the observation of Proxima b at close range from a multiplicity of viewpoints.  Swarm synchronization with state-of-the-art space-rated clocks would enable operational coherence if not actual phase coherence in the swarm optical communications.  Betavoltaic technology, which should be commercialized and space-rated in the next decade, can provide an adequate primary energy storage for these swarms.  The combination would thus enable data return rates orders of magnitude greater than possible from a single probe.
\end{abstract}




\maketitle

\section{Introduction} \label{SecIntro}
Per the direction of Breakthrough Starshot (BTS), we conducted a systems engineering study of the challenge of obtaining information from swarms of gram-scale probes across 4 light years from an interstellar target, hereinafter the planet Proxima b.  We find that this task is not impossible, and could be accomplished with six broad innovations: 

\clearpage
\thispagestyle{empty}
\begin{sidewaystable*}[ht]
  \centering
  \caption{Mission Traceability Matrix}
   \begin{tabular}{|c|c|c|c|c|}
\hline
Mission Goals & Problem &  Solution & Implications  \\
\hline
\hline
Surviving to Proxima b (1) & Dust impact ends mission with 1 spacecraft  & Fly many spacecraft & Swarm benefits described here\\ 
Surviving to Proxima b (2) & 20-MeV proton radiation at 0.2 c & Fly ``edge on.'' 
& Extra shielding in the rim \\
Assembling the Swarm  & Swarm is launched sequentially & Use ISM drag to cohere swarm & Ability to tack spacecraft as needed\\
Close Observation of Proxima b & Ephemeris errors of 10$^{5}$ km & 100 - 1000 Spacecraft swarm & Assembling the Swarm\\
Complete observation of Proxima b & Single spacecraft has only one vantage point
& Swarm has multiple vantage points & Enables transmission spectroscopy\\
Communicate with Earth during flyby & This will limit observations with that probe & Swarm elements can be sacrificed & Could have probes impact Proxima b\\
Interswarm Communication & 1000 - 10,000 km separation & Laser coms in spacecraft rim & Rim IR laser system\\ 
Regular communication with Earth & Need continuous power for decades & Betavoltaic power with $^{90}$Sr & 330 mg power system\\ 
Gigabytes of data return per year & Low photon rate with 1 spacecraft & ``Photon Coherent'' swarm laser coms & Swarm clock synchronization\\
\hline
\end{tabular}
\label{Table:Traceability}
\end{sidewaystable*}

\begin{table*}[ht]
  \centering
\begin{tabular}{|c|c|}
\hline
Swarm or Probe Parameter & Value  [units] \\
\hline
Transverse swarm diameter at flyby [km]                                             & 100,000   \\ 
Number of surviving probes at flyby, after 4-LY cruise                                             & 300   \\ 
Average probe spacing within swarm [km]                                             & 5700   \\ 
Individual probe diameter [mm]                                             & 4000   \\ 
Probe rim height [mm]                                                & 20  \\ 
Main disk thickness [mm]                                       & 10  \\ 
Mass budget: disk [mg] & 330 \\
Mass budget: betavoltaic battery and ultracapacitor pulsed storage [mg] & 330 \\
Mass budget: rim, inter-probe coms, computation and everything else [mg] & 340 \\
Overall input electrical power per probe, at flyby [mW] & 6 \\
Input electrical power to Swarm-Earth or inter-probe lasers, at flyby [mW] & 4 \\
Optical output power per probe, at flyby [mW] & 0.4 \\
Swarm-Earth communications wavelength source / as received red-shifted [nm]                     & 432 / 539 \\ 
Number probe-Earth communications apertures               & 247 \\ 
Inter-swarm (rim) communications wavelength [nm] & 12,000  \\ 
Number of rim transceivers per probe                           & 5   \\ 
\hline
\end{tabular}
\caption{Basic parameters of the proposed probe swarm. The aerographene metamaterial that forms the main probe body has a variable density, tailored for the particular mechanical requirements. A denser layer will support the drive-beam dielectric reflector while the middle layer will be very sparse. There would be thicker areas around the sensor/communications telescope array on the front face and around the betavoltaic, capacitor, and electronics layers to support them.}
\end{table*}
\begin{figure*}[ht!]
\includegraphics[scale=0.26]{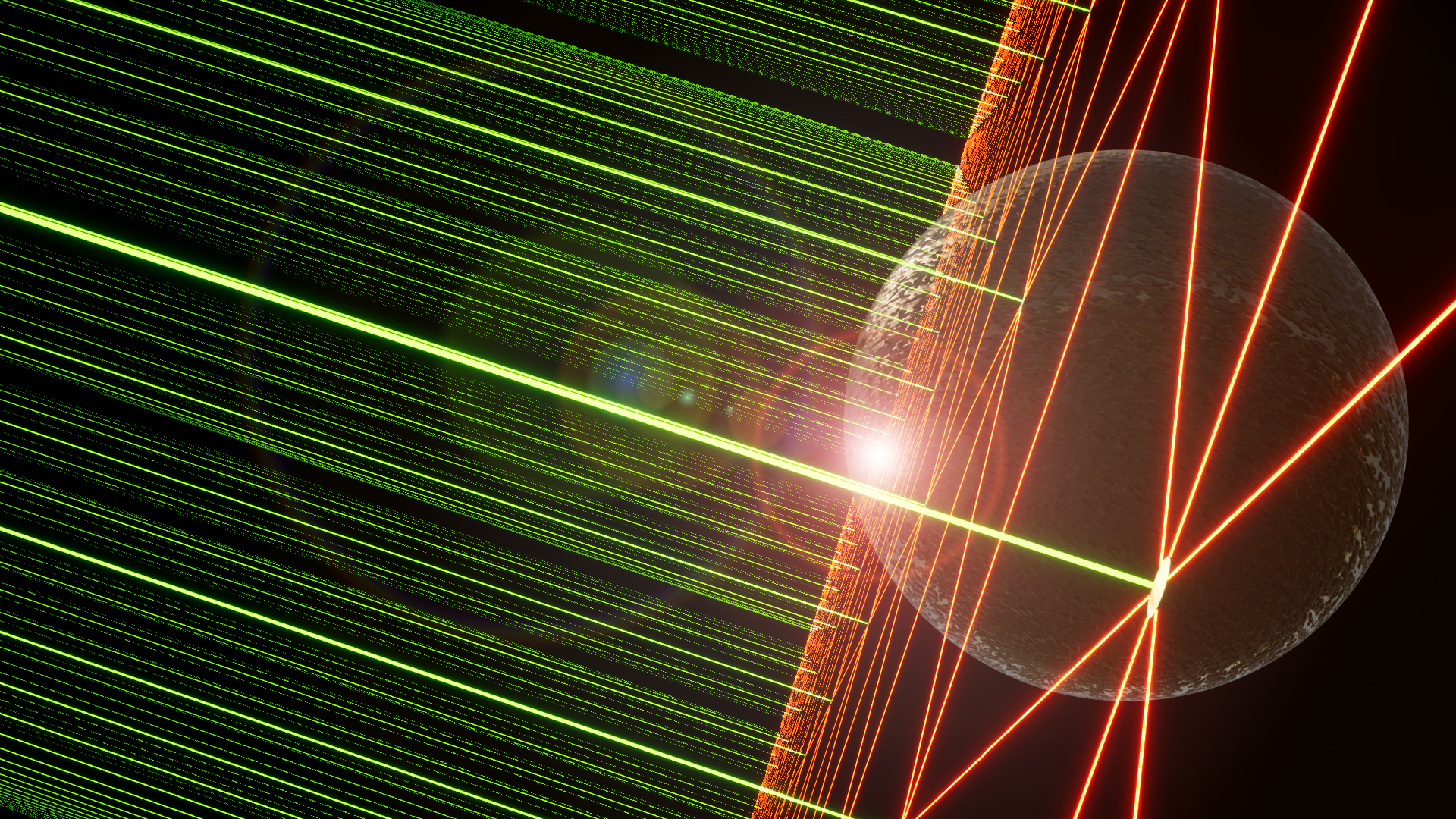}
\caption{Artist's impression of swarm passing by Proxima Centauri and Proxima b. The swarm's extent is $\sim$10 larger than the planet's, yet the $\sim$5000-km spacing is such that one or more probes will come close to or even impact the planet (flare on limb). It should be possible to do transmission spectroscopy with such swarms. Green 432/539-nm beams are coms to Earth; red 12,000-nm laser beacons are for intra-swarm probe-to-probe coms.  Conceptual artwork courtesy of Michel Lamontagne, P.Eng.}
\label{fig:Michelv2}
\end{figure*}

\begin{figure*}
\centering
\begin{subfigure}{1.0\textwidth}
\includegraphics[width=\textwidth]{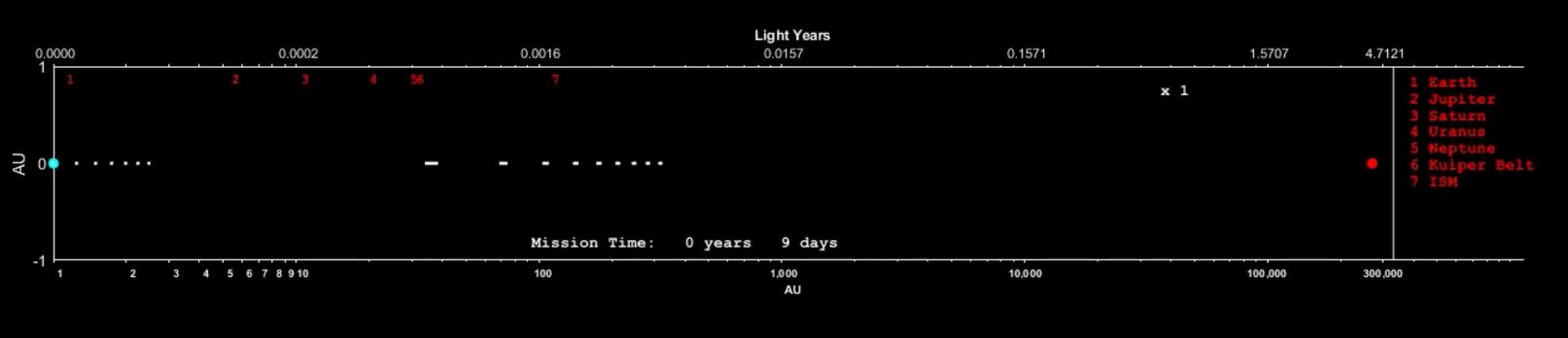}
\caption{A flotilla (sub-fleet) of probes (far left), individually fired at the maximum tempo of once per 9 minutes, departs Earth (blue) daily.  The planets pass in rapid succession.  Launched with the primary ToT technique, the individual probes draw closer to one another inside the flotilla, while the flotilla itself catches up with previously-launched flotillas exiting the outer Solar system (middle) $\sim$100 AU. For the animation go to https://www.youtube.com/watch?v=jMgfVMNxNQs  \cite{Break_vid}.  }
\end{subfigure}
\
\begin{subfigure}{1.0\textwidth}
\includegraphics[width=\textwidth]{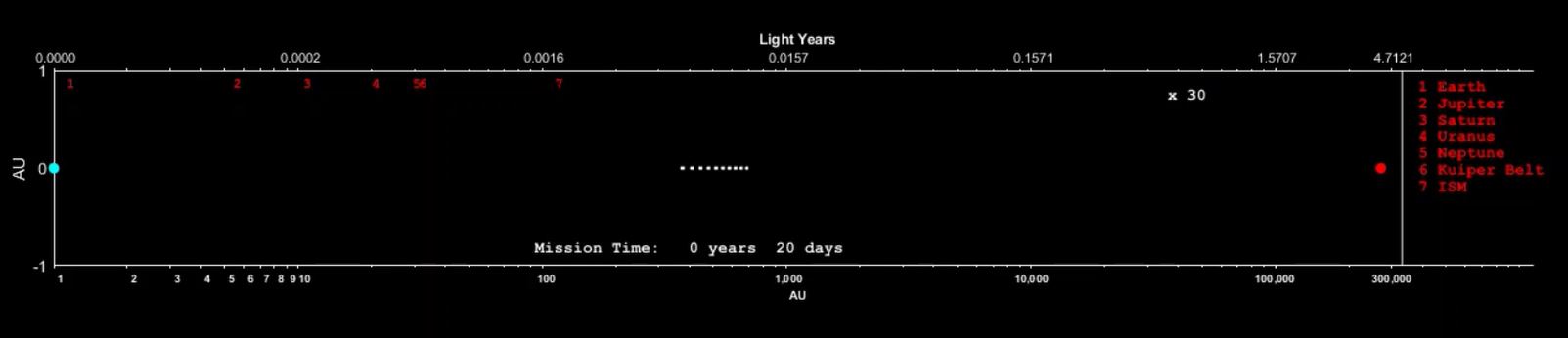}
\caption{Time sped up by a scale factor of 30.  The last flotilla launched draws closer to the earlier flotillas; the full fleet begins to coalesce (middle), now under both the primary ToT and secondary VoT techniques, beyond the Kuiper-Edgeworth Belt and entry into the Oort Cloud $\sim$1000--10,000 AU.}
\end{subfigure}
\
\begin{subfigure}{1.0\textwidth}
\includegraphics[width=\textwidth]{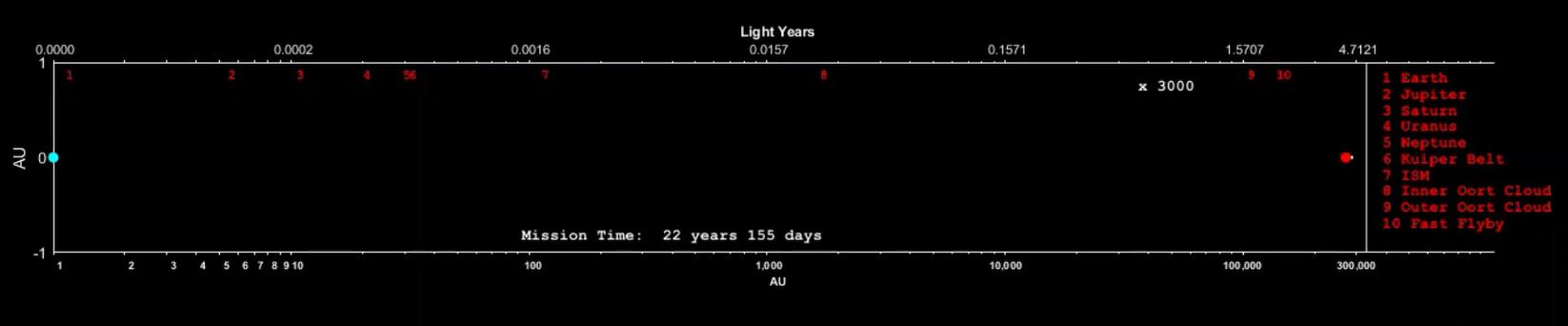}
\caption{Time sped up by a scale factor of 3000.  The fully-formed fleet, having reconfigured itself into a mesh network by autonomously applying the secondary VoT technique over decades of cruising, has exited Sol's Oort Cloud and flies by Proxima b at 0.2c. $\sim$300,000 AU.}
\label{fig:stills}
\end{subfigure}
\caption{Screenshots from CG animation at https://www.youtube.com/watch?v=jMgfVMNxNQs depicting launch and forming-up of the swarm. }
\end{figure*}

\begin{enumerate}
  \item Rather than individual 1-g spacecraft launched in isolation, instead send an ``operationally coherent'' (synchronized if not phase-coherent), swarm of 100s--1000s of spacecraft in order to simultaneously and feasibly transmit a reasonable number of signal photons to Earth.
  \item Forming effective swarms of order 100s--1000s of probes at planetary encounter with a gross ``time-on-target'' (ToT) technique, consisting of modulating the initial velocity of each probe by the launch laser such that the tail catches up with the head, in terms of their position relative to each other, not with Earth or Proxima b. 
  \item A finer ``velocity-on-target'' (VoT) technique based on controlled drag imparted by the interstellar medium (ISM) by altering the attitude of individual probes with respect to the ISM, thus keeping swarm together in relative and absolute position once formed.  Attitude adjustment is also necessary to minimize the extremely high radiation dose induced by traveling through the ISM at 0.2c. See Figure \ref{fig:stills}.  To minimize frontal area, hence dose and erosion, we anticipate flying edge-on most of the way, without rotation about the roll axis, hence leading and trailing edges are distinct, and configured differently, even though the probe is mostly symmetric.
  \item Intraswarm communication via optical means, to form up and then use the swarm as a signaling array. 
  \item Applying state-of-the-art (with frequency stability goal of 10$^{-19}$ at Earth \& 10$^{-13}$ at Proxima) optical clock metrology combined with a clever scheme of time- and frequency-bandpass filtering to improve data collection and signal-to-noise ratio (SNR) of data return to Earth. 
  \item A method onboard carrying stored energy sufficient for the entire mission, well-matched to its 2-decade duration, and in a compact form, i.e. at nuclear energy densities orders of magnitude greater than achievable with any known chemical method or in-flight electromagnetic or photonic method, to power computation and communication within the rigorous constraint of total payload not to exceed (NTE) one gram.  This particular betavoltaic source material, $^{90}$Sr, is many orders of magnitude cheaper than any other candidate, has a moderately high technology readiness level and could attain commercial-off-the-shelf (COTS) status within a decade with the right incentives. 
\end{enumerate}

 Finally, we have identified a fundamental operational unknown that must be solved well in advance: \textbf{accurately determining the orbital position of Proxima b at least 8.5 Earth years ($\sim$300 revolutions) ahead of launch}.  We find that this can be overcome at reasonable cost through the targeted use of gravitational microlensing observed by small telescopes in low Earth orbit (LEO).  

\begin{figure*}
\begin{subfigure}{1.0\textwidth}
\centering
\includegraphics[scale=1.0]{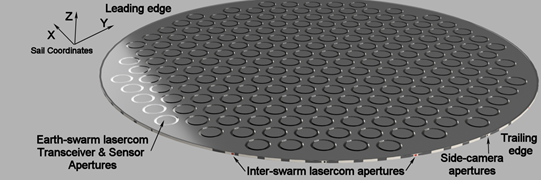}
\caption{Oblique view of the top/forward of a probe (side facing away from the launch laser) depicting array of phase-coherent apertures for sending data back to Earth, and optical transceivers in the rim for communication with each other.}
\label{fig:pauls}
\end{subfigure}
\begin{subfigure}{1.0\textwidth}
\centering
\includegraphics[scale=0.5]{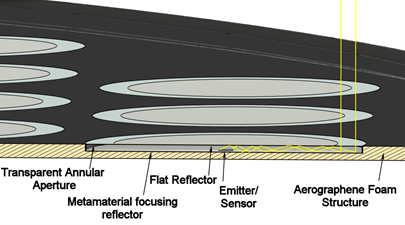}
\caption{Cross-sectional close-up view of one annular aperture in array of phase-coherent elements depicting ray trace from annular opening to sensor / emitter.}
\end{subfigure}
\begin{subfigure}{1.0\textwidth}
\centering
\includegraphics[scale=1.0]{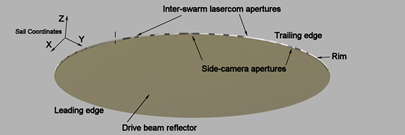}
\caption{Oblique view of bottom / aft of a probe (side facing the launch laser) depicting dielectric boost layer and and optical transceivers in the rim.}
\end{subfigure}
\begin{subfigure}{1.0\textwidth}
\centering
\includegraphics[scale=0.3]{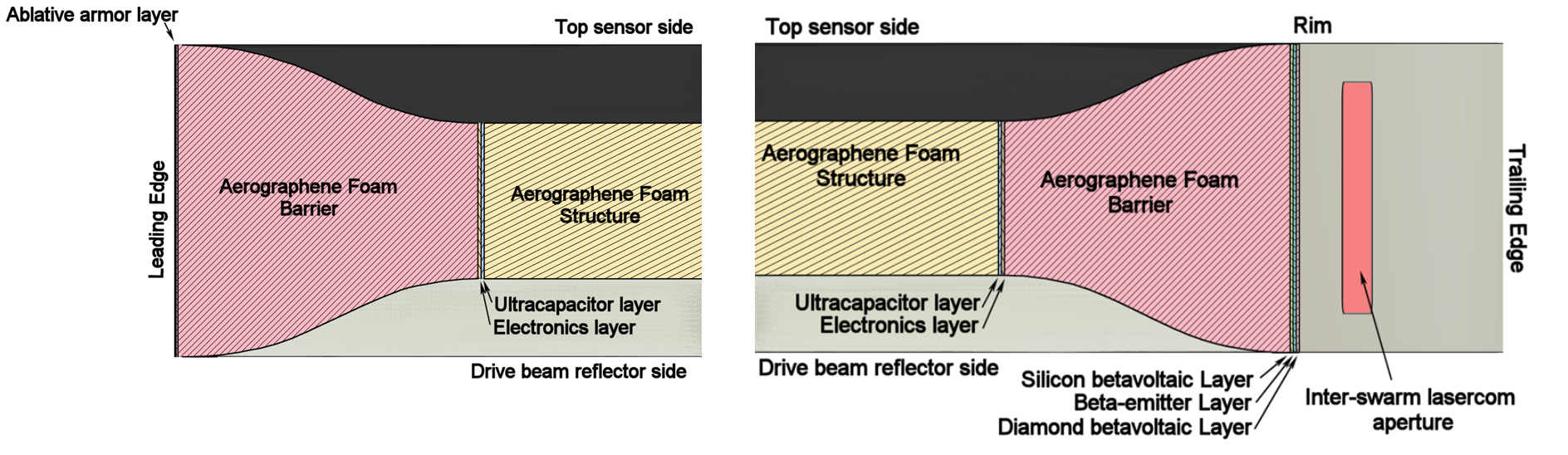}
\caption{Cross-sections of protective leading (to left) edge with sacrificial barrier, and instrumented trailing edge (to right) that contains the betavoltaic battery sandwiches and the probe-to-probe optical transceivers.  Note the electronic layer and ultracapacitor layer span the entire perimeter to assure connectivity.}
\end{subfigure}
\caption{Various views of Probe concept.}
\end{figure*}

We also have a few lesser innovations, including:
\begin{enumerate}
  \item A novel repurposing of the 100-gigawatt (GW) drive laser as an ``interstellar flashlight,'' illuminating objects in the path of the swarm (which we assess to be quite feasible).  This external light would aid in spotting small bodies at encounter, serve as a adjunct  monochromatic light source for spectroscopy by the fleet during flyby, as well as provide an additional marginal but possibly useful controlled source of illumination for photography of dark objects in Proxima system. 
  \item A novel structural concept, wherein most of the heart of the device (batteries, ultracapacitors, small-aperture inter-probe laser communication, computation) is concentrated in a 2-cm high thickened rim, while the central disk consists of a thin but large-aperture phase-coherent meta-material disk of flat optics similar to a fresnel lens for both imaging the target and communicating with Earth.  (Since valuable laser energy has been invested in accelerating both the sail and payload to 0.2c, we shall retain the dielectric layer on the aft face of the probe to serve as armor when the probe has to turn full-face on to the ISM in order to send and receive to Earth.)  The layout is reminiscent of a red corpuscle rather than the simple featureless flat disk which seems to be the default in the community. Refer to Figure \ref{fig:pauls}. 
\end{enumerate}

\section{Approach}

The broad goal is to send information from one or more gram-scale spacecraft across 4+ light years.  The default reference mission for the current BTS work is contained in \cite{Parkin_2018}.  Basic assigned constraints are: 0.2c cruise velocity, 200-nm sail thickness, 4.1-m sail diameter, 3.6-gram mass.

\subsection{Our Mission Vision}
Our swarming approach relies on launching a pipeline of hundreds of probes serially, maintaining continuous contact with each other and with Earth, and gradually, over the course of the cruise phase, reforming the long string into a flat lens-shaped mesh network $\sim$100,000 km across by the time of encounter.  Continual pinging of swarm by Earth with the boost laser will keep mission director continually apprised of the attrition rate due to collisions with microscopic interstellar objects  or other hazards. These spacecraft would be lower-mass and faster than the femtospacecraft of Project Andromeda \cite{Hein-et-al-2017-b}. Novel dynamic techniques described below would be utilized to cohere the swarm.

We note for the record that although all probes are assumed to be identical, implicitly in the community and explicitly in the baseline study, there is in fact no necessity for them to be ``cookie cutter'' copies, since the launch laser must be exquisitely tunable in the first place, capable of providing a boost tailored to every individual probe.  At minimum, probes can be configured and assigned for different operations while remaining dynamically identical, or they can be made truly heterogeneous wherein each probe could be rather different in form and function, if not overall mass and size.

\subsubsection{Time on Target}
``Time on target'' (ToT) means that either velocities, or paths, or both, are adjusted to bring multiple projectiles  on a target at the same time. For relativistic spacecraft the paths are largely fixed, so ToT relies on variation of velocity; in the remainder of this section velocities are assumed to be scalar quantities. 
Suppose two probes are fired at different times with \textbf{v}$_{1}$
being the velocity of the first probe, launched at time \textbf{T}. The target is distance $L$ away and it thus takes, in the Newtonian approximation, the first probe a duration, D, $=$ $\mathrm{L} / |\mathbf{v}_{1}|$ to arrive, which thus occurs at a time T + D.  Suppose that the second probe is launched a time \textbf{$\Delta$t} after the first, at a velocity \textbf{v$_{2}$} $=$  \textbf{v$_{1}$ + $\Delta$v }.  
The second probe’s velocity is adjusted by $\Delta$v so they both reach the target at the same time, albeit with different velocities. That can be solved, yielding $\Delta$ $v = v$ ($\Delta$ t / D).\\
If D = 20 years and $\Delta$t = 1 hour, then $\Delta$ $v =$0.3 km/s.
$\Delta$v scales linearly with $\Delta$t, so that 10 probes launched at longer 1-hour intervals would need, for example, a 3 km/s velocity difference between the head and tail.

\subsubsection{Time and Velocity on Target}

Time and Velocity on Target (VoT), adjusts both acceleration and velocity, and is the fundamental means of forming a coherent swarm in flight.  We achieve this by launching the last probes with higher speeds and using variable geometry to give them a higher drag (deceleration), so that they slow down to the speed of slower leading probes as they reach their location.\\
A string of probes relying on the ToT technique only could indeed form a swarm coincident with the Proxima Centauri system, or any other arbitrary point, albeit briefly.  But then absent any other forces it would quickly disperse afterwards.  Post-encounter dispersion of the swarm is highly undesirable, but can be eliminated with the VoT technique by changing the attitude of the spacecraft such that the leading edge points at an angle to the flight direction, increasing the drag induced by the ISM, and slowing the faster swarm members as they approach the slower ones.  Furthermore, this approach does not require substantial additional changes to the baseline BTS architecture.
The analytic approximation for exploiting the drag from the incoming  neutral hydrogen ISM  is \cite{hoang2017interaction}:
\begin{multline}
\frac{\Delta\mathrm{v}}{\mathrm{v}}\ \approx\ 
\left(\frac{2.5\ \times\ 10^{-6}}{M}\right)\ 
\left(\frac{\mathrm{N_{H}}}{10^{18}\ \mathrm{cm}^{-2} }\right)\ \\ 
\left(\frac{0.2\ c}{\mathrm{v}}\right)^{2.6}\ 
\left(\frac{\mathrm{A_{sail}}}{1\ \mathrm{m}^{2}}\right)\ 
\left(\frac{\mathrm{l}}{1 \mu m}\right)\
\label{eq:Hoang-drag}
\end{multline}

where $\Delta$V/v is the ratio between the velocity change due to the ISM and the initial launch velocity, $M$ the spacecraft mass, N$_{H}$ the column density of interstellar hydrogen between Earth and the target star, $v$ the initial spacecraft velocity, A$_{/mathrm{sail}}$ the sail surface area and l the sail thickness.  Using the analytic approximation and baseline dimensions (\textit{v} = 0.2c; \textbf{M} = 3.6 g; \textbf{d} = 4.1 m; \textbf{l} = 100 nm) from the Starshot RFP, we get ratios on the order of 10$^{-6}$.  It can be seen in the table that the launch tempo and thickness of aerographene are inverse linearly related quantities.\\

\begin{table*}
\begin{center}
\begin{tabular}{ |c | c | c | c | c| }
 \hline
Thickness of  & Spacecraft thickness
& Mass of   &  Total spacecraft
  & Velocity \\
aerographene layer   & with aerographene sheet&  aerographene layer  &   mass  &  Ratio  \\
\hline
100 nm & 200 nm & 2.4 $\times$ 10$^{-7}$ gm &  3.6 gm  & 10$^{-6}$ \\
1 $\mu$m & 1.1 $\mu$m & 0.0024 gm &  3.6 gm & 10$^{-5}$ \\
100 $\mu$m & 100.1 $\mu$m & 0.24 gm &  3.84 gm & 10$^{-3}$ \\
1 mm & 1 mm & 2.4 gm &  6 gm & 10$^{-2}$ \\
10 mm & 10 mm & 24 gm &  27.6 gm & 10$^{-1}$ \\
\hline 
\end{tabular}
\caption{Aerographene layers added on top of the spacecraft in flight direction with the same surface area as the spacecraft and their impact on spacecraft mass and generated interstellar hydrogen beam drag (v = 0.2c; M = 3.6 grams; d = 4.1 m.) The fifth column is the ratio of velocity lost from interaction with the ISM to initial launch velocity.  }
\label{table:data-solution}
\end{center}
\end{table*}

As before, suppose two probes are fired at different times with \textbf{v}$_{1}$ being the velocity of the first probe, launched at time T, and the second probe be launched a time $\Delta$t after the first, with a velocity difference of $\Delta$v. In this case, the target is a nominal point on the trajectory where the swarm can cohere; this formation is then retained for the rest of the mission. Suppose that this target point is mid-way to the target.  A point 2.12 light years from Earth would be reached at a time $\tau$ = 10.6 years into the mission for the first probe.  Assume that the imposed drag differential acceleration is \textbf{a}, and is constant over the first half of the mission. This can be solved, yielding

\begin{equation}
\mathrm{a}\ =\ -\frac{2\ \mathrm{v}_{1}\ \Delta\mathrm{t}}
{(\tau - \Delta\mathrm{t})^2 }
\label{eq:VOT-a}
\end{equation}
and 
\begin{equation}
\delta\mathrm{v}\ =\ \frac{2\ \mathrm{v}_{1}\ \Delta\mathrm{t}}
{(\tau - \Delta\mathrm{t})}  .
\label{eq:VOT-v}
\end{equation}

The operational objective is to dissipate a portion of the velocity of leading probes by continually adjusting their attitude hence aspect ratio and sectional density with respect to the oncoming ISM (edge on, fully face on, or something in between), in pitch and yaw axes, such that the hindmost members catch up with but do not overtake the leading members.\\
If the collisions with the ISM are elastic, then the reaction would also generate some useful cross-range velocity of order $\sim$1 km/sec, transverse to the direction of travel.  The VoT-Attitude Adjustment method could be initially be under Earth's control, but soon (due to communication latency well before hitting the Oort Cloud), it would have to become fully autonomous, i.e., under the control of individual probes and eventually that of the fleet, in effect a ``hive mind''.  With virtually no mass allowance for shielding, attitude adjustment is the only practical means to minimize the extreme radiation damage induced by traveling through the ISM at 0.2c.  Moreover, lacking the mass budget for mechanical gimbals or other means to point instruments, then controlling attitude and rate changes of the entire craft in pitch, yaw, roll, is the only practical way aim onboard sensors for intra-swarm communications, interstellar coms with Earth and imagery acquisition / distributed processing at encounter. \\
\underline{Magnetorquers.} In addition to interacting with the ISM, adjusting attitude with onboard ``magnetorquers'' (\textbf{$\tau$} $\sim$ \textit{I} $\times$ \textbf{B}) might be feasible, as is done for 1-kg Cubesats in low Earth orbit (LEO) right now with commercially-available space-rated products \cite{Magnetorquer_URL}.  
\begin{figure}
\centering
\includegraphics[scale=0.42]{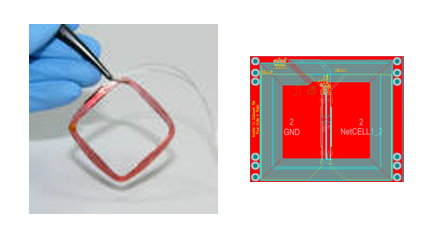}
\caption{Commercially available magnetorquers for Cubesats (left) and implemented as chips (right)}
\end{figure}  
However, applying this attitude adjustment technique to a 1-gram probe for BTS would require reducing the mass and size of presently available magnetorquers by many orders of magnitude. Assessing the feasibility of this method within the mass and time constraints of BTS will be left to the next phase of work.\\
\underline{Interaction with B-field in the ISM}.  If a probe accumulates an electric charge, q, then Lorentz force is expected to be a factor (\textbf{F} = \textit{q} (\textbf{\textit{v}} $\times$ \textbf{B}).  Although the magnetic fields in the ISM are thought to be very weak, the velocity \textit{v} would be very high, so the net effect may provide a basis for attitude control.  Assessing the feasibility of this method within the mass constraint will be left to the next phase of work.\\
\underline{Other Means of Interaction with ISM Assessed}.  
Given that the luminosity of Proxima Centauri is but $\sim$0.1\% of Sol's, in other words $\sim$1 W m$^{-2}$ at 1 AU, then photonic deceleration by that star alone would be ineffective.  For thoroughness, we also examined and modeled the stellar radiation of $\alpha$ Centauri A and B as a means of photonic deceleration, as has been proposed by \cite{heller2017deceleratio}.  However, the bright pair A/B are a fifth of a light-year away from Proxima, the deceleration would be infinitesimal.  We assess the photonic deceleration technique as insufficient to the BTS task.\\
\underline{MEMS trim tabs.}
Although we envision our conceptual probe as being entirely solid state, with no moving parts, a simple mechanism may become necessary in order to provide a rapid way to adjust attitude.   Consider the common aircraft ``trim tab'' $\sim$ but 1 cm square, actuated by microscopic electromechanical lever or other simple MEMS machine.  This is conceptually similar to the trim tabs proposed by R.Angel in 2006 for the school of trillions of 1-gram flyers near the Sun-Earth Lagrange 1 (SEL1) point.  \cite{Angel2006}  
\begin{figure*}
\centering
\includegraphics[scale=0.5]{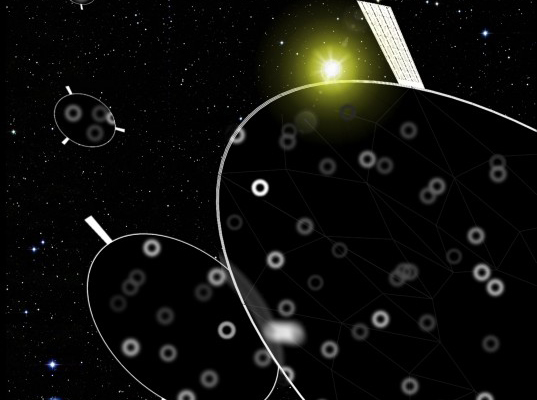}
\caption{1-gram, 70-cm glass diffraction gratings with photonic pressure trim tabs. \cite{Angel2006}}
\end{figure*}
A few such trim tabs could be spaced symmetrically in pairs around the Probe's rim to provide a means of dynamic control.  When it is raised, it interacts with the ISM to generate an asymmetric torque on the probe; to halt the torque, the tab is lowered; to stop the rotation caused by the torque, the tab opposite the first tab is raised then lowered.  Of course, the ``drag-brakes'' will be withdrawn after the probes reach the target zone, so that the differential acceleration \textbf{a} goes to zero and the velocities thereafter remain the same. 

\subsubsection{Optical Communications by a Swarm}
Transverse laser links between probes, which we have modeled, would inform an even finer degree of autonomous station-keeping.  But a swarm of probes has a coordination problem after it is launched---at first, its members will not know where the other probes are.  We have developed preliminary protocols to get a coherent swarm, or a well behaved synchronized one at any rate, and in the process, discover how large a swarm might reasonably be.  Result: A swarm of order $\sim$100,000 km diameter can be established in deep space!  This will take time, but time is something we have a lot of. 
``Self-knowledge'' will occur in several distinct phases:

\begin{enumerate}
\item \underline{Discovery} Like fireflies, members have to find and establish communications with each other.  We assume that the swarm have been placed into a cluster by the drive process, which is now over.  Assume each probe is slowly rotating (order 0.5 rpm) about its line of flight, defined as X-axis, per left-hand side of the figure.  
\begin{figure*}
\centering
\includegraphics[scale=0.4]{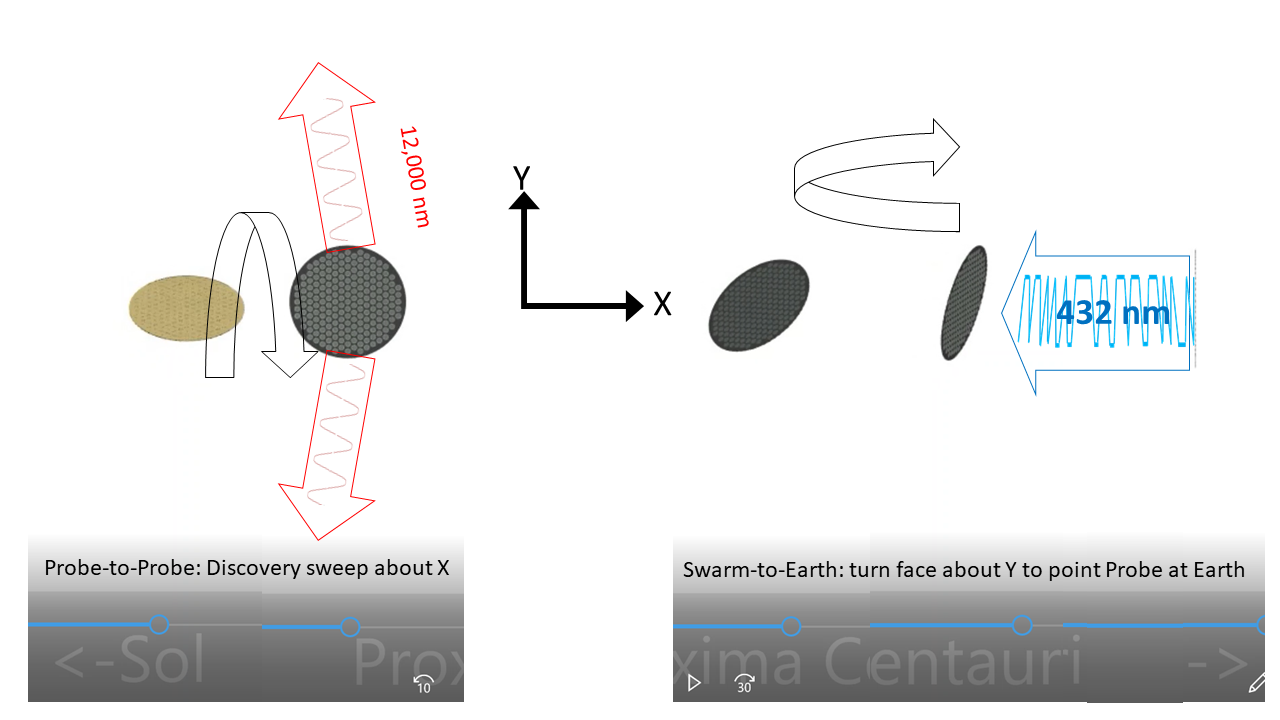}
\caption{Rotations for Intra-Swarm Probe-to-Probe Com, and for Swarm-to-Earth Com.  Note the rotation alternately presents the forward coms then the gold dielectric boost layer.}
\label{fig:ProbeXYRotations}
\end{figure*}
For the animation go to https://www.youtube.com/watch?v=iD3S5Qj5sR4  \cite{Dance_vid}).  The probe has multiple 1-mW pulsed optical transceivers around its rim with 20-mm flat lenses and quantum dot lasers, for sending and receiving to other members, and a mutual view period (due to rotation) of order 15 msec, then each probe should should be detectable by its closest neighbor out to $\sim$5,700 km.  
\item \underline{Probes as Beacons} 
During the initial link-up phase, the probes will have to find each other to form a swarm. Their approximate locations can be uplinked from Earth, but the probes will have to locate each other for communications using internal resources. For this, we assume that the intra-swarm communications receive optics can be adjusted for a much wider field of view of up to 45 degrees and that the transmit beam is formed into a thin vertical fan, several arc-seconds wide by 45 degrees high. Thus, as the probe rotates during cruise (again, the optics are along the bottom half of the rim, over a span of 160 degrees) each probe transmits a pulsed beacon while simultaneously scanning for its neighbors. 
The output power, again, is calculated by assuming that the total generated power of 4 mW is stored in a super-capacitor and available for laser pulsing.  Any one probe should be able to detect another probe at a range of $\sim$1 million km.
\begin{figure*}
\centering
\includegraphics[scale=0.6]{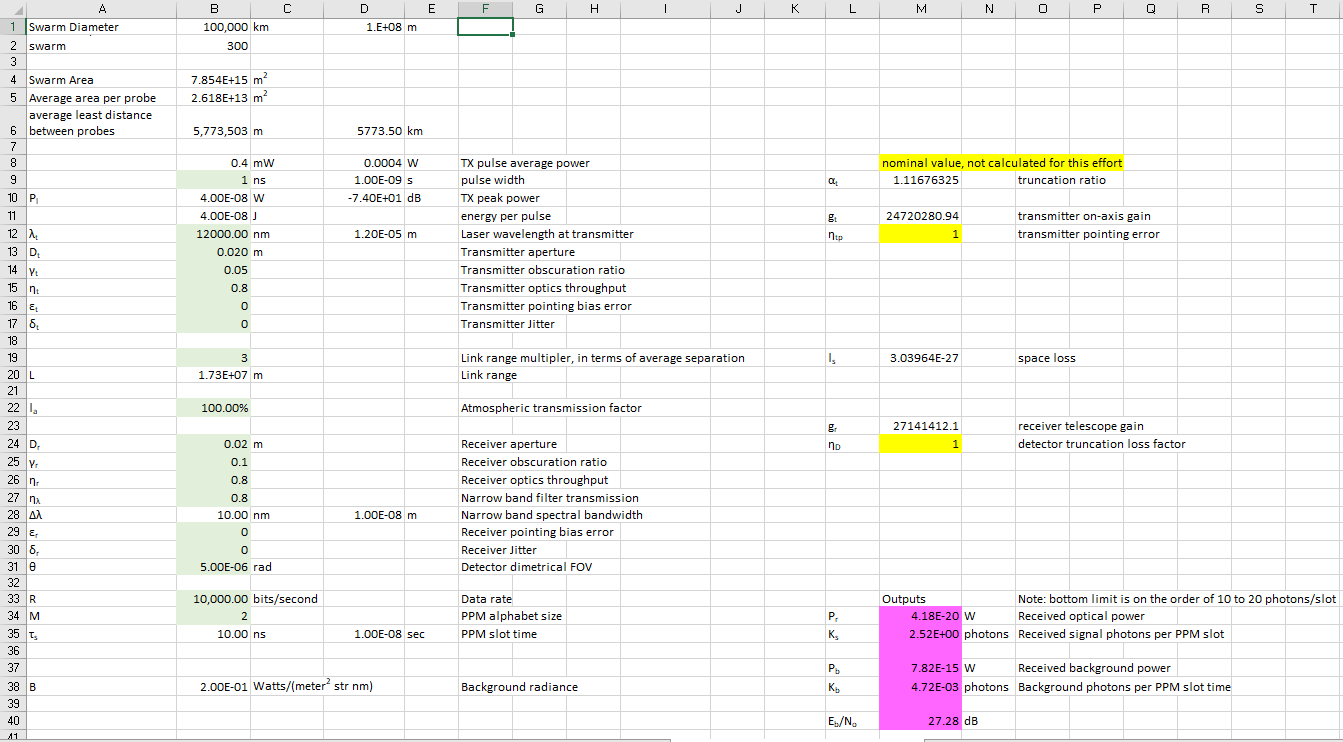}
\caption{Worksheet for Probe-to-Probe Communication.}
\end{figure*}
\item \underline{Convergence}. Setting up a mesh network using a modification of Mobile Ad Hoc Networking is based on Manet.  Low-bit identity tokens can be passed between member pairs.   Eventually, a distance vector ``map'' of visibility of other nodes can be iteratively established by each node.  This map might include angles as well as actual distance and SNR.  Nodes can then efficiently transmit distance vector information to other nodes such that the swarm becomes aware of itself. The goal of convergence is to establish mesh communications between all nodes in the swarm.  The goal of alignment is to optimize that mesh communication.  \\
\item \underline{Alignment} is still a work in progress. \\ 
\item \underline{Intra-Swarm Probe-to-Probe Link Budget}\\
The communications link budget for intra-swarm communications is calculated using the link-budget method.  The equations are the same as for the swarm-Earth budget below, but only one transmitter and one receiver are used. The transmit/receive apertures are 2-cm diameter flat-optic devices spread around the back, or trailing, half of the outer rim of the probe, which flies edge-on most of the time to minimize radiation dose and erosion by particle flux induced by the ISM at 0.2c. We assume that the optics are electronically steerable, so that high gain may be achieved. Again, the beam is collimated to the degree allowed by the diffraction limit of the aperture. \\
Instead of being visible, the intra-fleet transmit laser wavelength is $\sim$12,000 nm, or long-wave infrared (IR).  This greatly reduces the path loss, drawing less power.\\ 
We don’t know what direction the side-looking rim ports on the probe will be pointing at any given time, but we can be sure the Centauri system stars will not be directly in the field of view.  Therefore, the background is assumed to be the general sky noise used by astronomy.  The receivers are equivalent to avalanche photo diodes (APD). 
Given a swarm of 300 probes with a diameter of 100,000 km, and assuming that the probes are evenly spaced, the average distance between probes will be about 5000 km. If we allow the maximum path to be triple this, so that each probe can talk to as many neighbors as possible, then with a bit rate of 10 kHz we get 2.5 photons per time slot and a SNR of 27 dB. 

\item \underline{Swarm-to-Earth Communication} using Photon Counting and Link Budget Methods\\
2-symbol Pulse Position Modulation (2-PPM), as shown in figure \ref{fig:PPM} is widely used in optical communications links.  2-PPM uses synchronous time slots, with two adjacent time slots for each bit. A ``0'' value is sent with a pulse in slot 1 and no pulse in slot 2; and \textit{vice versa} to send a ``1'' value.  During processing the first slot is subtracted from the second slot; the background will cancel out.  A positive result indicates a ``1'' value, a negative indicates a ``0''. Pulses are 1 ns in duration and the integration slots are 10 ns in length. Symbols are transmitted at 10 Hz.  This technique also has the desirable effect of greatly lowering the unwanted background noise, since the integration time is very short. \\(Higher order PPM symbols may be used.  For instance 4-PPM encodes two bits, with four discrete values, into each symbol using a laser pulse in one of four time slots.  However, for simplicity, we did not consider these in this paper.)  \\
Given a final velocity of 0.2c, a transmitted wavelength of 432 nm (blue), is Doppler-shifted to 539 nm (green) when received at Earth.  
\begin{figure}
\centering
\includegraphics[scale=0.3]{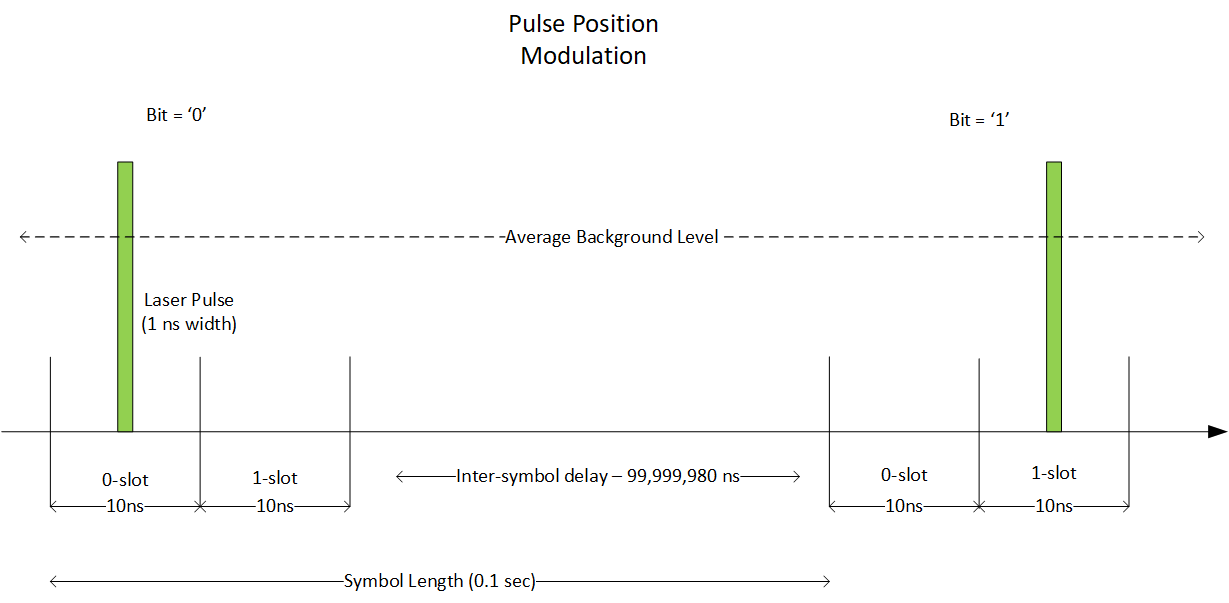}
\caption{Pulse Position Modulation}
\label{fig:PPM}
\end{figure}  \\
The first table below depicts a method based on estimating direct photon transmission from the swarm to Earth.  The photon flux for the laser pulse is calculated and summed for the swarm. We assume that the optics collimate the output beam to the degree allowed by the Airy disk of the optics; for simplicity we assume that the beam flux is uniform across the beam's width. Once the beam reaches Earth, we factor in the atmospheric transmission, based on the average value for the laser wavelength, and calculate the number of photons captured by the entire array, within the 2-PPM slot time.  We also sum the background flux from both the Milky Way galaxy, which is directly behind the probe swarm in the field of view, taking into account the Airy disk angle of the receiving telescope, and the three stars in the Centauri system, which will also be in the field of view. By calculating the number of photons arriving during the 2-PPM time slot, we can estimate the noise levels on top of the signal.  Applying this method, we end up with 724 signal photons captured by the array per pulse, from an original count per pulse of 2.6x10$^{16}$ with an average background level of less than one noise photon per time slot. 
\begin{figure*}
\centering
\includegraphics[scale=0.45]{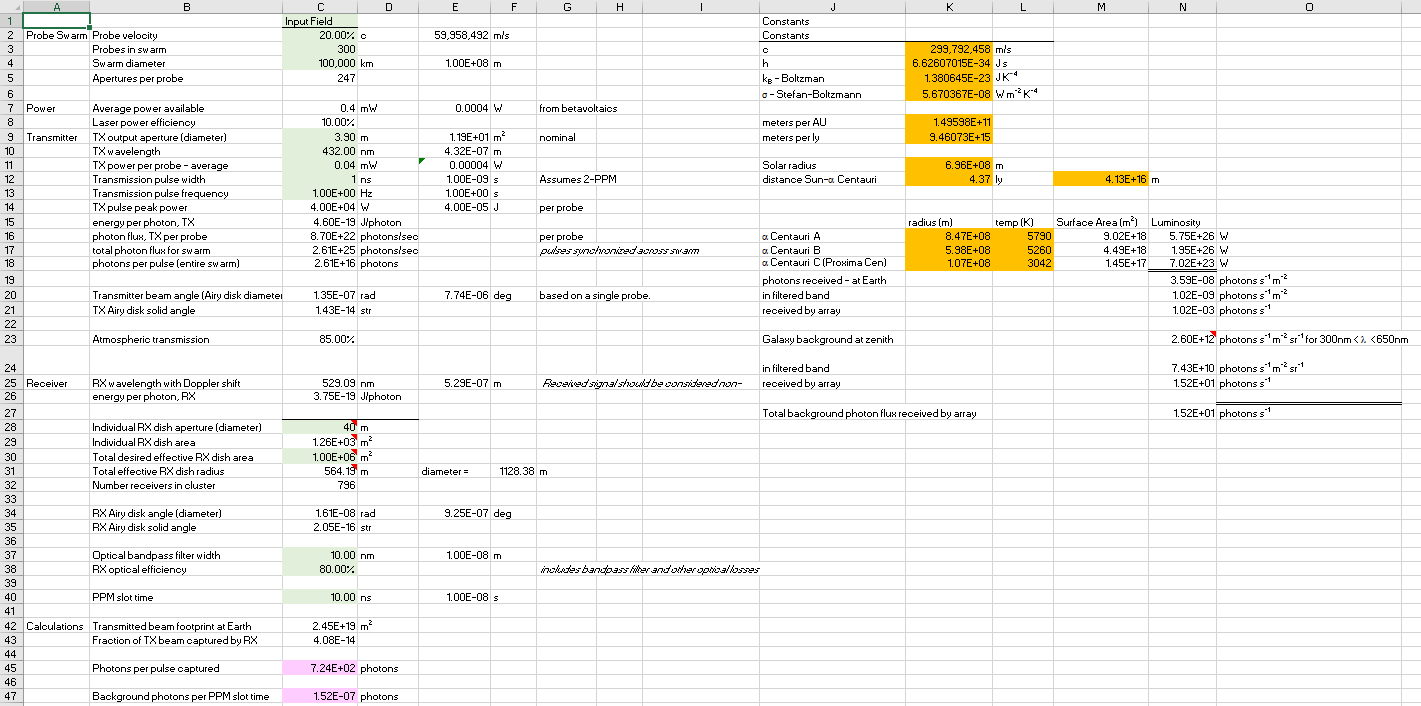}
\caption{Worksheet for Swarm-to-Earth Communication, Photon Counting Method.}
\label{fig:Earth-photon-count-spreadsheet}
\end{figure*}

The second method is the traditional link-budget algorithm used in both optical and RF communications.\\ 
\begin{equation}
\label{Power}
P_{r} = P_{t} G_{t} G_{r} L_{s} L_{a} \eta_ {pt} \eta_{t} \eta_{r} ,    
\end{equation}
where P$_{t}$ is the transmitted power; \\
G$_{t}$ is the transmitter gain; \\
G$_{r}$ is the receiver gain; \\
L$_{s}$ is the path loss; \\
L$_{a}$ is the atmospheric loss; \\
$\eta_{pt}$ is the pointing efficiency; \\
$\eta_{t}$ is the transmitter efficiency and
$\eta_{r}$ is the receiver efficiency. \\

The transmitter and receiver gain are calculated from the diffraction limits of the telescope apertures for the laser bean wavelength and are expressed relative to an omnidirectional ``antenna''. \cite{Marshall1986ReceivedOP, Yuen2022, Wang-et-al-2014-a} . 
For simplicity and because we do not have reliable estimates, in this method, we assume that the transceiver and the receivers both have perfect pointing accuracy.  Using this method we get 386 received photons per PPM time slot and again, an average background level of less than one noise photon per slot.

\begin{figure*}
\centering
\includegraphics[scale=0.5]{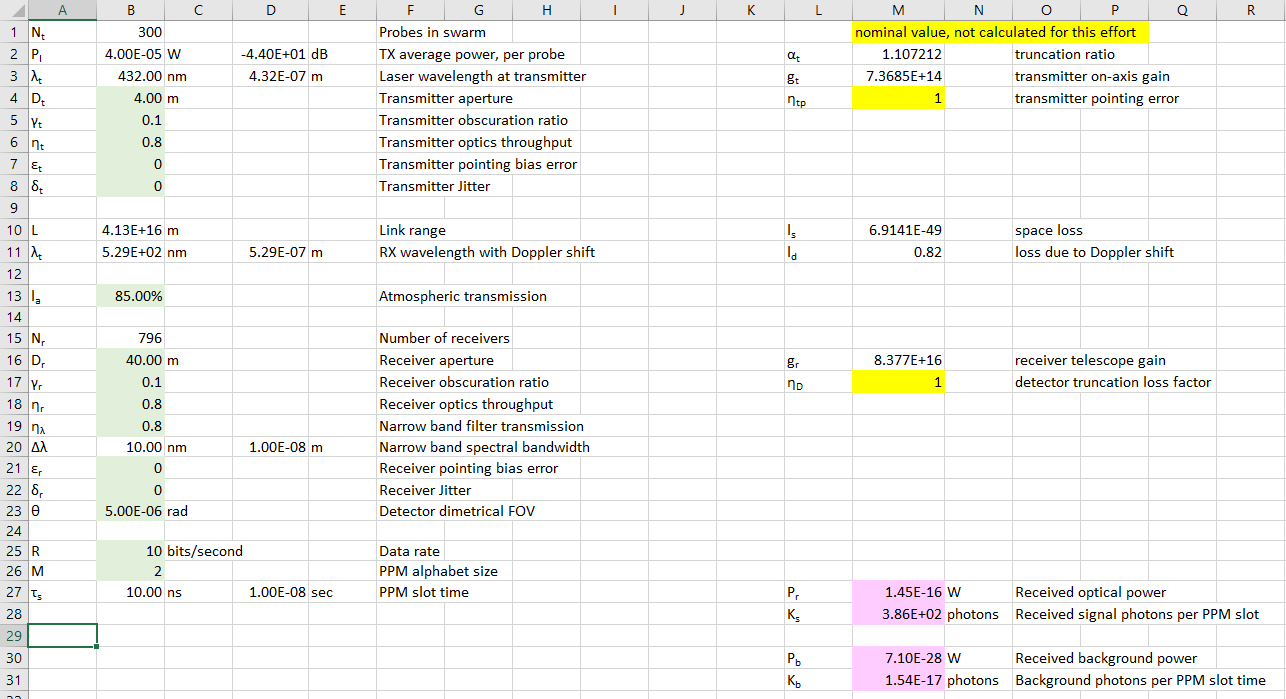}
\caption{Worksheet for Swarm-to-Earth Communication, Link Budget Method.}
\label{fig:Earth-link-budget-spreadsheet}
\end{figure*}
For each probe, the electrical power is provided by a betavoltaic battery generating a total of $\sim$10 mW at launch, decaying to 6 mW at the time of flyby, of which 4 mW is partitioned to the main laser com system.  This is converted into 0.4 mW of optical power, assuming today's mere 10 percent electric-to-photonic conversion efficiency.  (We expect this efficiency to improve dramatically over the next few decades.)  Electricity can be stored in a rapid-discharge ultracapacitor.  By concentrating power into 1-ns pulses, the average power of each laser pulse is 4x10$^{4}$W, containing 40 micro-joules per pulse. 
\end{enumerate}

\subsubsection{Onboard Clock Metrology and Science Return, or The Role of Really Good Clocks in Swarming Proxima}

\begin{figure}[ht]
\centering
\includegraphics[scale=0.7]{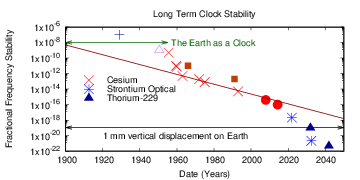}
\caption{Long Term Trends in Clock Frequency Stability. The red icons are for various types of Cesium clocks and frequency standards used in official timekeeping, blue icons are real or predicted optical atomic clocks, and the black icons predicted Thorium nuclear clocks. The solid horizontal line is the gravitational red shift (-1.095 $\times$ 10$^{-19}$) caused by a 1 mm altitude difference on the surface of the Earth. As it is very difficult to model or measure absolute vertical motions on the Earth to better than $\sim$1 mm, when clocks improve beyond this limit it is likely that the best time standards will have to migrate into space.}
\end{figure}

Better timekeeping has always been a disruptive technology throughout human history \cite{Kennedy2021}.  However, the rate of technological progress since World War II has been exceptional, literally a decade of frequency stability per decade on the calendar.  Furthermore, the differential cost trends are almost as attractive as Moore's Law---while $\sim$\$2000 buys an instrument that can measure voltage or current to five places, the same money buys an astounding \textbf{\textit{twelve}} places of precision for measuring time.\\

Therefore, our innovation is to use advances in optical clocks, mode-locked optical lasers, and network protocols to enable a swarm of widely separated small spacecraft or small flotillas of such to behave as a single distributed entity.  Optical frequency and reliable picosecond timing, synchronized between Earth and Proxima b, is what underpins the capability for useful data return despite the seemingly low source power, very large space loss and low signal-to-noise ratio.  While quantum metrology would limit error in interferometry to 1/N (N is number of photons received) vs. classical proportionality (1/$\sqrt(N))$, the state of the art is not quite there yet. \\

\begin{figure}[ht]
\centering
\includegraphics[scale=0.5]{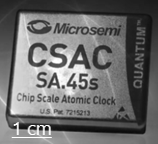}
\caption{Space-rated radiation-hardened chip-scale atomic clock.  Credit: Microsemi Product Catalog: https://microsemi.com/product-directory/embedded-clocks-frequency-references/5207-space-csac.}
\label{CSAC}
\end{figure}

Nevertheless, state-of-the-art chip-scale atomic clocks (CSAC), existence now and likely to be space-rated in the future, can enable the simple synchronization of optical pulses amongst numerous individual emitters -- which we dub ``operational coherence'' -- if not the proper ``phase coherence'' required for true ``interferometry in reverse''.  With a very limited energy budget, synchronization squeezes the same signal photons into a smaller transmission window, temporarily increasing the brightness of the signal relative to unavoidable background noise.  With sufficiently narrow reception bins on Earth, this means data rates much higher than from a single probe with the same mass.  Such a swarm-based mesh network would also provide information about the target from a multiplicity of look-angles and distances, \textit{information that would not be available in any other way}.\\
Another intriguing very-long-term possibility is an atomic-scale clock built around a single-atom oscillator.  \cite{Hannah&Brown2007}  Although we assess that such devices would not be ready soon enough for BTS, improving a swarm's synchronization to beyond picosecond levels would provide true phase coherence, enabling an entire fleet to act as a single transmitter with vast aperture, which in turn would focus many orders of magnitude more photons to receivers in the Sol system than individual gram-scale spacecraft could possibly manage on their own, even if they were individually phase-coherent. In the more distant future, beyond the time frame of this BTS mission, a truly phase-coherent swarm would both require and enable autonomous position-navigation-timing (PNT) down to 100-m precision for receiving optical timing pulses from the Earth.\\

\subsubsection{Signal Processing at Earth}
\begin{figure}
\centering
\includegraphics[scale=0.3]{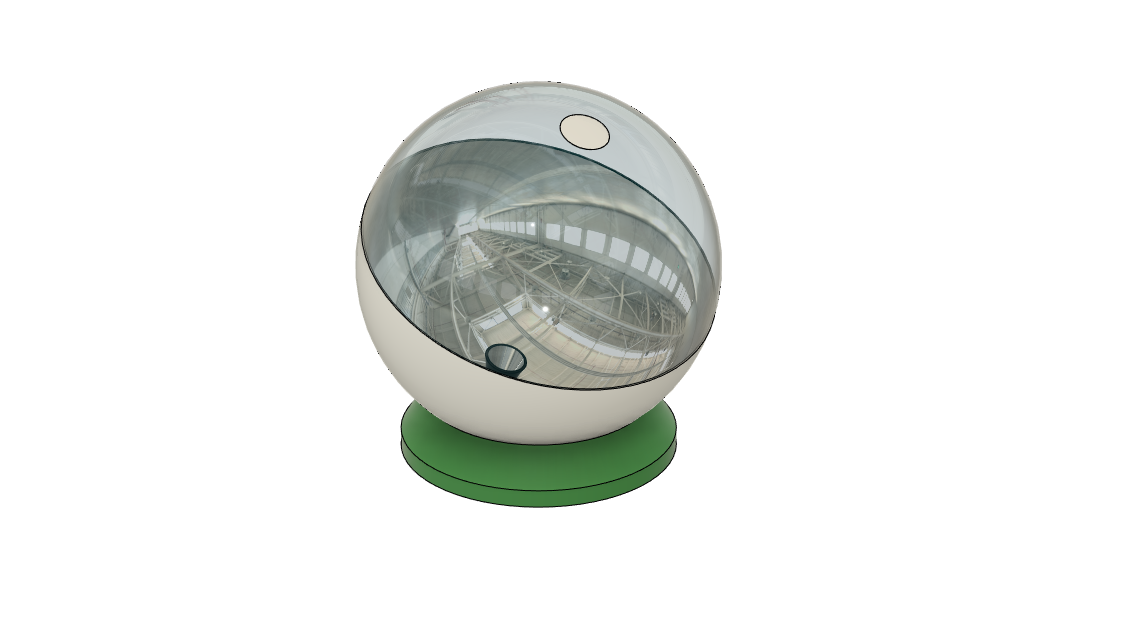}
\caption{A conceptual receiver implemented as a large inflatable sphere, similar to widely used inflatable antenna domes; the upper half is transparent, the lower half is silvered to form a half-sphere mirror. At the top is a secondary mirror which sends the light down into a cone-shaped accumulator which gathers it into the receiver in the base.  The optical signals would be received and converted to electrical signals – most probably with APDs at each station and combined electrically at a central processing facility.  Each bucket has a 10-nm wide band-pass filter, centered on the Doppler-shifted received laser frequency. This could be made narrower, but since the probes will be maneuvering and slowing in order to meet up and form the swarm, and there will be some deceleration on the whole swarm due to drag induced by the ISM, there will be some uncertainty in the exact wavelength of the received signal.}
\label{fig:LightBucket}
\end{figure}

We assume a large array of 796 ``light buckets'' (Figure \ref{fig:LightBucket}) on Earth, summing to 1 km$^{2}$ (the size of a city in aggregate), and $\sim$10$^{19}$ photons in each signal chirp. These pulses are assumed to occur once a minute and to be operationally-coherent or synchronized from the whole fleet, so that a few hundred 539-nm signal photons can be expected to arrive at the receiver on Earth during particular narrow reception windows in time and wavelength. This transmission mode requires  good clocks and good PNT at the source, and also time synchronization with the Earth.\\

Hence the importance of good clocks to the entire scheme.  Therefore the final piece of our approach is to apply clever time- and frequency-bandpass filtering back on Earth to maximize the chance of seeing these photons, which should stand out against Proxima Centauri's weak UV flux.  Even without the cleverness, the amount of computation that can be dedicated to signal processing at home is essentially unlimited.  While electronic receiver design is not covered in this paper, with the essentially unlimited post-processing that we anticipate to be available by 2050, a photon signal-to-noise ratio (SNR) of 1:1, or 0 dB, is sufficient to receive a signal with a reasonable bit error rate. For the actual embodiment in a real Probe, we assume that additional error-detection-correction schemes, such as Reed-Solomon or Turbo codes, will be used.  We are indeed fortunate.\\

\subsection{Science Goals and Means During Boost, Cruise, Flyby and Post-Encounter}

\subsubsection{Boost Phase}
The operational, engineering, and scientific intelligence gathering can commence immediately upon launch, since the probes will have been boosted to the highest velocities of the entire mission (they only slow down after launch, at greater or lesser rates), and the density of collision hazards (mostly dust in Sol system) is by far the highest it will ever be. 
 
\subsubsection{En Route}
\begin{figure}
\centering
\includegraphics[scale=0.42]{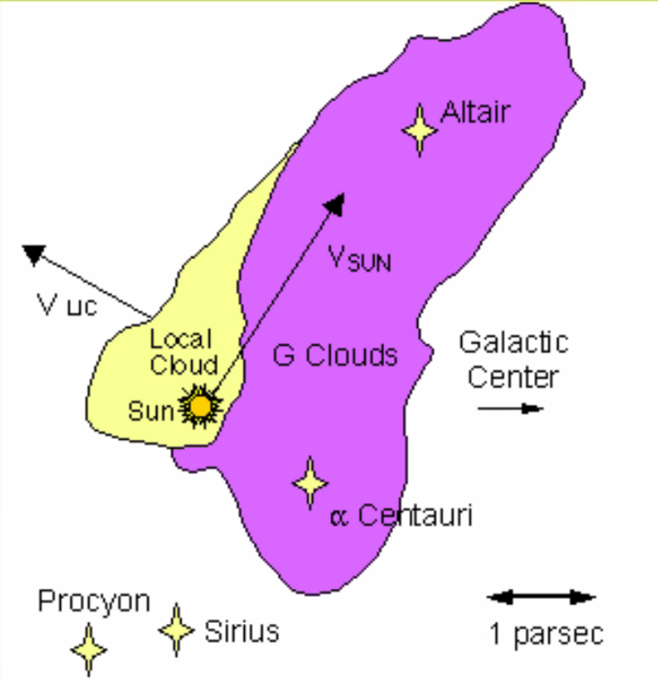}
\caption{Local clouds in the Interstellar Medium (ISM) from the North Galactic Pole, based on inversion of Doppler shifted absorption lines \cite{Linksy-et-al-2019-a}. }
\label{fig:ISM1}
\end{figure}

\begin{figure}[ht]
\hspace{0.5cm}
\includegraphics[scale=0.75]{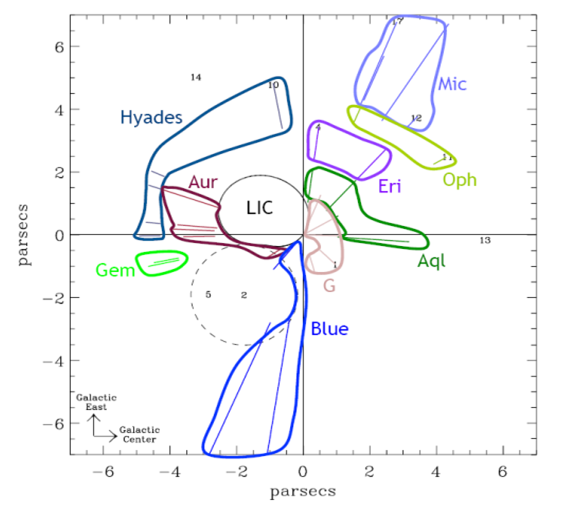}
\caption{Another view of local clouds in the ISM, in the Galactic X-Y plane as seen from the North Galactic Pole \cite{Linksy-et-al-2022-a}.  Our Solar System might be on the boundary of ``G'', or a few thousand AU inside or outside that boundary (a distance the size of a period here on those plots).  Resolving this ambiguity is an important goal of initial deep precursor missions, but is also crucial for planning of the BTS mission itself, as drag from these clouds will be an important tool for swarm coherence.}
\label{fig:ISM2}
\end{figure}

We can glean much technical information en route valuable to many fields of science both practical and theoretical.  Currently, our direct \textit{in situ} observations of the ``nearby'' ISM are limited to Voyager spacecraft at 100 AU, which have not fully departed our heliopause.  Indirect estimates rely on absorption lines in direction of nearby stars.  This means we only have data along lines of sight.  If there is no star in some direction, there is no measurement in that direction!  The nearest cloud is the Local Interstellar Cloud (LIC),  but it is not even clear if our solar system  is in that cloud or the G Cloud, which contains the Alpha Centauri system \cite{Linksy-et-al-2022-a}.  We might be on the boundary of G cloud, or a few thousand AU inside it, or away from it (see Figures \ref{fig:ISM1} \& \ref{fig:ISM2}).

\begin{enumerate}
  \item \underline{Pathfinding in the Sol-to-Centauri ISM}.   Recording the attrition rate of these pioneering pathfinding probes due to collisions with ISM bigger than single atoms or other hazards will both inform this mission's operation and also provide the first high-fidelity map of the entire tube of ISM all the way between Sol and Proxima, thus providing much basic  astronomical and stellar cartographic intelligence to the scientific community and the world at large long before arrival.  
  \item \underline{Chronometric geodesy}.  Accurate time synchronization between state-of-the-art clocks on Earth (whose size is unconstrained by spaceflight) tiny clocks aboard the fleet will provide important clues about gravity waves of such long wavelength they cannot be observed on or near Earth.  Scaling the known performance of state-of-the-art clocks in existence on Earth now would support determination of a spacecraft's position at Proxima to an accuracy of 100 meters, an astounding thought.  Perhaps a 10$^{-21}$ frequency stability based on transitions inside the atomic nucleus, which are even faster than optical transitions in the electron cloud, will be attained by the time of the fleet's launch, which is the best attainable on Earth.  For better than 10$^{-21}$, even the best clocks will have to move into space. 
 \end{enumerate} 
 
\subsubsection{Encounter}
\begin{enumerate}
\item \underline{Approach Geomtery}.  We want to approach the planet on its day side, obviously, in order to reveal macroscopic surface features such as bodies of water, ice, continents.  This approach also provides the best lighting geometry for finding moons, if any.  Therefore swarm must pass the star first (but not too close!), which sets a ti 
Refer to Figure \ref{fig:Flyby}. 
\begin{figure*}[ht]
\hspace{-0.5cm}
\includegraphics[scale=0.6]{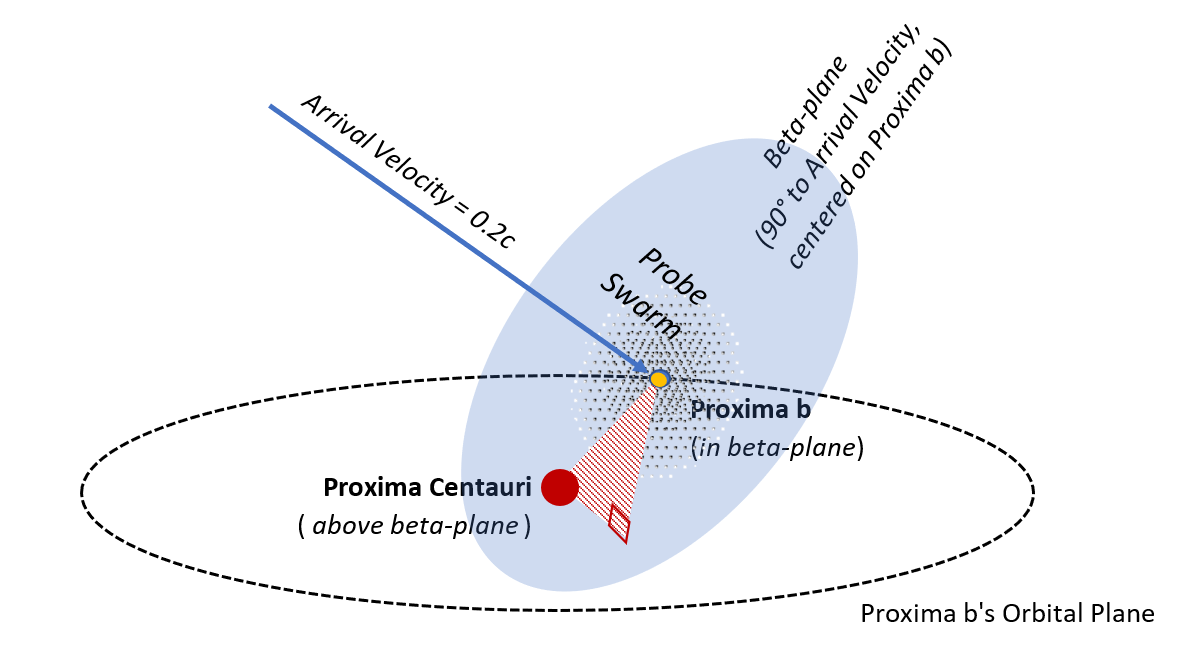}
\caption{	
	Geometry of swarm's encounter with Proxima b. The Beta-plane is the plane orthogonal to the velocity vector of the probe ''at infinity'' as it approaches the planet; in this example the star is above (before) the Beta-plane. To ensure that the elements of the swarm pass near the target, the probe-swarm is a disk oriented perpendicular to the velocity vector and extended enough to cover the expected transverse uncertainty in the probe-Proxima b ephemeris.}
\label{fig:Flyby}
\end{figure*}  
\item \underline{Gibbous Aspect}.  We want a gibbous aspect so we can see the terminator, which is where life would be if it exists anywhere on this world that probably is tidally locked (we should know for sure by the time this mission flies). The way to obtain a gibbous aspect is to arrange the launch time 20 years beforehand so as to arrive when Proxima b is at a moderate phase angle from Proxima Centauri, which provides a reasonably large fraction of sunlit hemisphere and extent of terminator, as shown, while not getting too close to the star.  (See Figure \ref{fig:Flyby}).  We have noted in this work that the inclination of Proxima Centauri's invariable plane with respect to that of Sol's, as well as the other orbital parameters specific to Proxima b such as the inclination of its ecliptic, its epoch, eccentricity, periastron, are not known today to the necessary accuracy if they are even known at all.  Resolving this basic lack of understanding about the Proximan system is an operational challenge that \textbf{must} be overcome decades before the fleet launches. \\ 
\item \underline{Impact Spectroscopy}.  Although the distribution pattern of the probes in the fleet would be approximately hex-close-packing for maximal communication efficiency, over the course of 20 years of flight, the mesh could gradually draw together and densify itself around a point expected to intersect the planet.  The distance of one probe to another at the fleet's center could be as little as a few thousand kilometers, in which case, one or more 1-gram probes may be expected to impact the upper atmosphere, if one exists, or the surface, it it does not.  However, lack of an atmosphere does not prove lack of life---Proxima b may be an ice world like Europa.  Each impact which would release $\sim$90 terajoules or $\sim$22 kilotonnes.  (Being equivalent to the yield of an old-fashioned ``atomic'' bomb, that would be one helluva flashbulb.)  Since a neutral hydrogen in the ISM releases 20 MeV when it hits the probe, the heavier elements in the probe can be expected to release $\sim$1 GeV per atom, which is far more than merely ionizing, this much energy will disrupt a nucleus.  We note that Earth experiences such blasts in the upper atmosphere at least annually with no ill effects.  The flash would be visible to nearly the entire fleet, and yield important spectroscopic data about the composition of Proxima's atmosphere.  It would also serve as an unambiguous timestamp.  If Proxima b is determined by some other method to host life then this scenario must be avoided. 
\item \underline{Transmission Spectroscopy}.  Upon arrival we should able to do transmission spectroscopy between pairs of spacecraft at Proxima b, because some members of the swarm will pass behind Proxima b as seen from the Earth, due to inherent transverse motions of Earth and Sol system with respect to Proxima b and Proxima Centauri.  Immediately after flyby of Proxima b by most of the swarm, we want to look back from the night side at the sunset of Proxima Centauri on Proxima b's limb through the thin ring of atmosphere, if it exists.  (At the same time, we might be able to spot cities if they exist and are artificially lighted.)   Furthermore, with sufficiently good timekeeping and absolute position measurement the drive laser could be utilized at that time to backlight Proxima b in monochromatic light which would be very useful for spectroscopy / spectrometry.  This would be a big deal scientifically, on the order of Webb’s main missions (a ``Flagship'' mission), or roughly equivalent to what the Starshade is.  (Being able to see each other fore and aft is also why the swarm must be lens-shaped, not be a completely flat disk, in order to have some longitudinal dimension along the line of flight.)
 \end{enumerate} 

\subsubsection{Post-Encounter}
\begin{figure}[ht]
\includegraphics[scale=0.68]{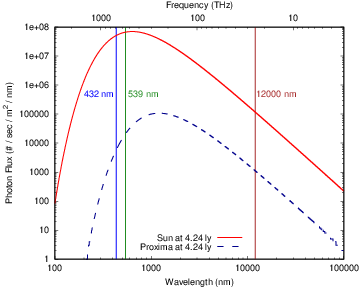}
\caption{Photon fluxes, in terms of photons m$^{-2}$ $\times$ s$^{-1}$ nm$^{-1}$ to and from Proxima Centauri 4.24 light years from each star, with superimposed wavelengths of laser coms: probe-to-probe, and swarm-to-Earth, as sent and received.  Flux from Sol at 1 $\mu$m is $\sim$500 $\times$ the quiescent Proxima flux.  Note that Proxima background emission drops off steeply below 1000 nm wavelength and would be very small at 432-nm source transmission or even the red-shifted 539-nm received signal.  For realistic smallsat apertures and spectral resolution, within a 1-ns-duration time gate, the laser photon flux from the target would be very much greater that the flux from the star itself (excepting stellar flares increasing brightness by $\sim$68, but are almost entirely emitted on specific spectral lines).}
\label{fig:PhotoFluxes}
\end{figure}

.  We assume that the probes require about 1 year to send back the data collected during the flyby of the $\alpha$ Centauri system.  Assuming a ``light bucket'' on or near the Earth that is in aggregate the size of a city 1 km$^{2}$, and order 100 joules per pulse once a minute from the fleet (containing $\sim$10$^{19}$ 432-nm photons), with good position-navigation-timing (PNT) at both ends, then with clever time and spectral frequency bandpass filtering, and sufficiently narrow reception windows, a few hundred photons can be expected to arrive to a 1 km$^{2}$ receiver on Earth with the pulse system described in Figures \ref{fig:Earth-link-budget-spreadsheet} and \ref{fig:Earth-photon-count-spreadsheet}.  These photons, red-shifted to 539-nm on arrival, should stand out against Proxima Centauri's very weak near-UV flux (see Figure \ref{fig:PhotoFluxes}).  Again, even absent the cleverness, the amount of computation that can be dedicated to signal processing back on Earth is essentially unlimited, which is fortunate.\\

\subsection{Predicting Proxima b, or Necessary Astrometry}.   
Since the goal is to fly this swarm past Proxima b, we need to predict the position of Proxima b to order 10,000 km at least 8.6 years before flyby (the time it takes for information between the swarm and Earth to complete one round-trip at the speed of light).  Since four years is roughly 10$^{8}$ seconds, predicting the position of Proxima b to 10$^{8}$ m (our swarm diameter) means determining the planet's velocity to order 1 m/s, and its angular position to $\sim$0.1 microradians.  A good optical read from a terrestrial telescope yields a position accuracy of order 4 x 10$^{9}$ meters so obtaining a velocity error of order 1 m/s implies a time baseline of order 100 years, which is too long for BTS. So, either one must go to space with a really big telescope, or use other techniques.\\
Although we already have Proxima b’s period (11.68 days), we need to determine its line of nodes, eccentricity, inclination and epoch, and also its perturbations by the other planets in the system. At the time of flyby, the most recent Earth update will be at least 8.5 years old.  The Proxima b orbit state will need to be propagated over at least that interval to predict its position, and that prediction needs to be accuracy to the order of the swarm diameter.  That implies an orbital velocity error $\lesssim$0.3 m/s at the time of the last update (or less than one part in 10$^{-5}$, and knowledge of the semi-major axis at that time to $\sim$8.5 km or better. \\
In addition, the star's ephemeris requires a position accuracy (both transverse and radial) of 10,000 km, requiring knowledge of both the  parallax and angular position (as seen from Earth) of $\sim$40 microarcseconds at arrival, requiring determination of the star's proper motion and parallax to one part in 10$^{5}$ at launch.  We think that this accuracy could be provided with a small spacecraft in Earth orbit tracking Proxima microlensing events (by matching the lens-source relative velocity and thus significantly lengthening the duration of the lensing events) and thus obtaining fine details of the motion of Proxima b.  But that’s a story for another paper. 

\subsection{Implications of Our Approach for BTS}
Our approach inevitably has profound implications for some aspects of the overall BTS architecture, particularly related to launch operations, however we have come to believe that the objective cannot be achieved in any other way if the principal constraints of ``gram-scale spacecraft'' and ``data return within a human lifetime'' are retained.\\

Our operational modeling shows that even firing just one probe per day from Earth for a year, it is possible to assemble a functional swarm of 100s--1000s of probes at Proxima b, noting that the Starshot baseline is 9 minutes per shot.  Larger swarms (10$^{4}$ to 10$^{5}$) would provide greater capability due to more eyes, more power and brightness, and more computation at encounter, as well as better odds of survival after 20 years transit through the ISM, would be possible with a faster launch cadence.  In addition, a larger swarm would both support a significantly higher bit rate (data return), and also provide the ability for some probes to observe their brethren near and even behind the planetAccoracy of time perspective impossible to gain with just one probe, or from Earth).

\section{Probe Concepts}   
\subsection{Material}
\begin{figure}
\centering
\includegraphics[scale=0.7]{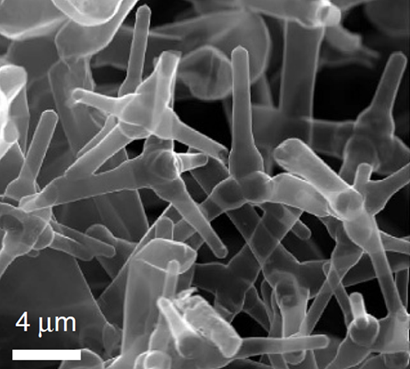}
\caption{Microphotograph of graphite tetrapod (an areographene material.}
\label{fig-graphene}
\end{figure}

We propose to significantly increase the thickness of the sail by utilizing extremely low-density materials to increase the collision cross section of the sail vis-à-vis the oncoming hydrogen flux.  Candidate materials could be aerographene \citep{shah2022synthesis} (density: 0.16 kg m$^{-3}$) and aerographite \citep{mecklenburg2012aerographite} (density: 0.18 kg m$^{-3}$) \citep{behera2021advanced}. To illustrate the superb performance of this material, a 1-mm thick aerographene structure would have an area-to-mass ratio of order of 10$^{-4}$ kg m$^{-2}$.  The ratio for a 1-micrometer thick structure would be $10^{-7}$ kg m$^{-2}$.  Due to this exceptionally low sectional low density, the performance of aerographene for the purpose of deceleration is about $10^{4}$ better than for Mylar.  Both materials are completely opaque with an absorptivity of 1.  Also, both materials have been synthesized \citep{mecklenburg2012aerographite} in the laboratory.  No principal roadblocks seem to exist towards mass production. Both materials share many similarities but for simplicity, we will focus on aerographite in this work.\\
Current structures based on this material are of order 10s--100s of $\mu$m \citep{meija2017nanomechanics}.  Further reducing the thickness to order 10s--100s of nanometers should, in principle, be feasible, as the wall thickness of the tetrapods in Figure \ref{fig-graphene} is on the order of 10 nm.  The use of smaller tetrapods as sacrificial material on which the graphite is deposited should be possible. \citep{meija2017nanomechanics}.  The tetrapod size and shape were based on the t-ZnO and t-AG sacrificial material used for depositing the graphite on those shapes. There do not seem to be principal obstacles to shrinking the tetrapod size to sub-µm to generate a µm-scale porous structure.  While the synthesis seems feasible, the main question is whether the mechanical properties of such a thin, porous structure can satisfy the requirements for an interstellar mission. \\

\subsection{Layout and Features}
The 10,000-gravity launch condition is so extreme that concentrations of mass (``mascons'') must be absolutely avoided.  Therefore the thickness of the probe must be as uniform as possible.  The layout that results from this constraint is reminiscent of a red corpuscle rather than the simple flat disk which seems to be the default in the community.  
The central disk consists of a large-diameter (4m) phase-coherent array of fresnel-like flat optics  consisting of 247 smaller 25-cm annular apertures arranged in hex-close-packing (hcp, like a honeycomb).  Each smaller 25-cm element consists of an annular aperture above an optical well containing a central sensor.  Because they are part of a common monolithic structure, each of those 247 elements knows exactly where it is in relation to the others, to within a few nanometers, which is what makes phase coherence by the big optical element possible.  The diffraction limit for the beam is based on the diameter of the whole array.  The phased array is for both imaging the target and emitting coherent photons for communicating back to Earth.  Because it contains 247 separate elements, the array will possess a great deal of redundancy. Therefore the array will tolerate a high rate of attrition due the microscopic holes that will inevitably be formed by impacts with neutral hydrogen or helium nuclei at those times when the probe must cruise face-on to the ISM.  Although photonic power would be lost with each failure of an annular element, down to a certain level, information would not be lost, like missing dots in an LED traffic light.\\

Assuming that a third of the 1-gram payload mass is partitioned to this one item, the entire probe must still be extremely thin, no greater than 28 microns if it were a solid plate (which it is not) but several hundred microns if a sandwich of aerographene between meta-material films.  \\
Another third of the probe's mass is partitioned to the nuclear battery.\\
Everything else comprises the remaining third.\\

\begin{table}[ht]
\caption{Mass Partition of the Probe}
\label{masspart}
\begin{tabular}{|l|r|}
\hline
Main optical disk (\%)                     & 33 \\ 
Betavoltaic cells and Ultracapacitors (\%) & 33 \\ 
Rim and Everything else (\%)                       & 34 \\ \hline
\end{tabular}
\end{table}

Other than the main optical disk, the functional heart of a probe (batteries, ultracapacitors, 2-cm-aperture lasers for inter-probe communication, computation and other housekeeping electronics) is concentrated the 160-degree long trailing edge of a 2-cm high thickened raised rim perpendicular to the disk.  \\
Since the probes will fly edge-on without rotation on the roll axis in order to minimize the radiation dose, the leading edge of the probe will consist of a 2-cm thick layer of aerogel faced with a thin sheet diamond to serve as an sacrificial barrier to absorb impacts from the ISM and dissipate the energy via evaporation.  No electronics are located any closer than 2 cm deep to the outer surface of this barrier, nor do any instrument ports penetrate the surface.  The barrier extends around 200 degrees of the probe's circumference.  No functionality is contained within this buffer except the single layer of electronics buried 2-cm deep under the face of the leading edge which helps assure overall connectivity.

\subsection{Construction and Manufacturing}
The 1-gram constraint is so extreme that it precludes traditional frameworks built up of subassemblies or conventionally-packaged chips with pins on circuit boards or radiation shielding.  Even existing surface-mount technology such as in Raspberry-Pi devices would not be good enough to avoid mascons.  Therefore, we assess that probes must be fabricated with a combination of methods:\\

\begin{enumerate}
\item additive manufacturing such as stereolithography at macro scale and the multi-pass processes used at microscale now to manufacture flash memory (which can contain $>$200 distinct layers); \\
\item direct ion implantation;\\
\item wafer-scale integration - state-of-the-art today is a single chip 30 cm across containing trillions of transistors. The ``30-cm'' limit is defined by the maximum size of silicon ``log'' that the semiconductor industry is willing to produce today, which is an order of magnitude smaller than the probe.\\
\end{enumerate}

However, these methods would have to occur at \\
\begin{enumerate} 
\item near atomic-scale, perhaps with atomic force microscopes (AFM), and \\ 
\item at near-atomic precision, and\\
\item at very high speeds or massively parallel operations due to the astronomical number of atoms contained even with a 1-gram spacecraft. \\ 
\item Furthermore, the overall extent of the workpiece (up to 4.1 meters diameter per baseline parameters) is 9 orders of magnitude larger than $\sim$1-nanometer size of individual features that would be directly laid down.  
\item An additional complication would be working with radioactive materials which must be directly placed without packaging or shielding due to the mass constraint.  The facility that can accomplish all this would only vaguely resemble a modern chip ``fab''.\\
 \end{enumerate} 
 
 \subsection{Power}
In the course of this work, it became apparent that the communications link budget must be grounded in a rational power budget, which is why this team invested the time to work out a defensible power system.  Given the extreme design constraint on the spacecraft, allowing just one gram for everything except the dielectric layer during launch, every speck of mass must ``earn its keep'' every day over the duration of the entire voyage, not just during the brief encounter.  We have concluded that as high as mass fraction as possible for stored energy  must be brought along.  For comparison, the Voyager probes of two generations ago has 15 percent of their total mass partitioned to the $^{238}$Pu power system; we believe a 30 percent set-aside for nuclear energy storage / power generation would be prudent for the BTS probe.  However---a subtle but key point---\textbf{it would be pointless to bring any stored mass-energy that would last any longer than the transmission phase.} \\
\subsubsection{Photovoltaic Options}
Even though triple-junction silicon cells have achieved photoelectric conversion efficiencies $\ge$40 percent in the laboratory, based on the criterion above, we immediately eliminated photovoltaic from consideration as its effective duty cycle would be less than 0.0001 percent (10s of minutes out of a 10-million-minute-long mission), which conflicts with the principle ``earn your keep every day''.   Furthermore, the luminosity of Proxima Centauri, which peaks in the infrared \ref{fig:PhotoFluxes}, is but $\sim$0.1\% of Sol's, i.e. $\sim$1 W m$^{-2}$ at 1 AU, or $\sim$200W for less than a minute during flyby.  At the time of this writing, it is not prudent to count on generating photovoltaic power with as-yet unknown infrared-optimized semiconductors of unknown efficacy.
\subsubsection{Other Electromagnetic Options}
If they worked at all, ``e-sails'' and ``mag-sails'' would have a 100-percent duty cycle during the entire voyage , but were nevertheless eliminated from consideration as their technological readiness is far less than the isotopic power option described below.
\subsubsection{Radio-Isotope Thermoelectric Generators}
The loss of mass in the fission of one uranium-235 nucleus yields $\sim$200 million electron-volts (MeV), i.e., $\sim$1 MeV per nucleon, which is 7 orders of magnitude greater than the energy of chemical transitions taking place in an electron shell.  This intensity is due to the transition being mediated by the strong force.  Lesser transitions such as radioactive decay (alpha, beta, gamma emissions) governed by the electroweak force at 10s to 100s of kiloelectron-volts (keV) per nucleon are 2-3 orders of magnitude smaller, yet due to their nuclear origin are still thousands of times greater than achievable with any known chemical method or in-flight electromagnetic or photonic method.  Furthermore, time constants typical of middle-range radioactive decay is well matched to the necessary duration (decades) of interstellar flight even at relativistic speeds. \\
\textbf{$^{238}$Pu} Highly miniaturized 150-mg plutonium-238 radioisotope thermoelectric generators (RTG) were discussed in early days of BTS.  However, RTGs suffer from two fundamental physical constraints relevant to this BTS mission.  Thermal process requires a heat source and a cold sink.  As an object gets smaller, its surface-to-volume ratio increases in inverse proportion to the length reduction, according to the Square-Cube Principle, which means proportionately more radiating area per unit mass to dissipate heat.  This is great for the cold sink.  The more profound challenge is the fundamental limits to miniaturizing anything whose function is based maintaining continuously high temperature in the heat source.  According to the Stefan-Boltzmann Law, radiation goes as T to the fourth power.  Very small hot objects cool off very quickly, hence why metal sparks and meteors behave the way we observe them to, and whereas much of the Earth is still liquid 4 billion years after its accretion.  This is a strong curve to be on the wrong side of.\\  
The second flaw is the relatively poor efficiency of thermionic conversion, in the range of a few percent.  Furthermore, there is a logistical challenge, in that the demand for $^{238}$Pu for space missions already far exceeds the supply.  Even though $^{238}$Pu is not fissionable hence no nuclear weapons potential, which should exclude it from the class of ``special nuclear material'' (SNM), it is a strong alpha emitter, which triggers extreme regulatory precautions all the same.  Even though a fleet of 1000 probes would require but a few grams of $^{238}$Pu, distribution in any amount is tightly controlled.  Finally, according to the committee at the National Academy of Sciences (NAS), which studied the matter of supply, no one has even proposed studying the miniaturization of $^{238}$Pu RTGs.  Given the generally glacial pace of progress that is peculiar to the nuclear field, laboratory-scale experiments that commenced today to miniaturize RTGs would not yield results for many decades. \\
\textbf{$^{90}$Sr} RTGs fueled with strontium-90 (in the form of a sintered ceramic, strontium titanate, SrTiO$_3$) were widely deployed in the Soviet Union for remote off-grid technical applications.  Strontium-90 is far more available (roughly 1 kilotonne in the world today), far cheaper (pennies per curie versus hundreds of dollars per curie), and far less hazardous to work with than plutonium-238, yet still has enough radioactivity to get quite hot depending on concentration and freshness.    
\subsubsection{Nuclear batteries}
\begin{figure}
\centering
\includegraphics[scale=0.4]{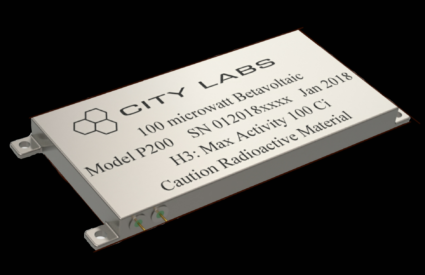}
\caption{NanoTritium$^{TM}$ betavoltaic nuclear battery fueled by 100 curies of tritium, which generates 100 microwatts (0.1 milliwatts).  Picture courtesy of City Labs, Inc., 12491 SW 134th Court, Suite 23, Miami, FL 33186. }
\end{figure}

\begin{enumerate}
\item \underline{Past and Existing Betavoltaic Cells}.\\  
\textbf{$^{147}Pm$}  In the 1970s, the ``Betacel'', a prototype betavoltaic battery based on promethium-147 was successfully designed and built to provide years of power for surgically implanted pacemakers.  However, promethium-146, a sister isotope that emits a hard gamma ray, was inextricably mixed with the desired isotope.  The two were impossible to mass-separate given their mere 1-amu mass difference, therefore most of the weight and volume of the device was for necessary shielding.  Furthermore, the idea of a radioactive implant faced some resistance in the medical market.  The invention of the lithium battery at just about the same time doomed the Betacel.\\

Tritium decays to helium-3 in a single step by emitting a single $\beta$$^{-}$ ray (which is an electron, as opposed to an $\alpha$ particle which is a much more massive ionized helium nucleus).  The sole commercial-off-the-shelf (COTS) betavoltaic power source available today is the NanoTritium$^{TM}$ battery, fueled with 100 curies (10 milligram) of tritium, generating an initial 100 microwatts (0.1 milliwatt) per unit at the moment of manufacture.  \\  
This COTS technology would be immediately applicable to less demanding precursor missions within the Solar system.  Moreover, it appears that \textit{``Betavoltaics are like Jupiter!''} meaning that commercially-available products appear to maintain a constant form factor regardless of a three-order-of-magnitude range in power output.  This in turn suggests that an nearly arbitrary number of curies (from 225 millicuries in early tritium cells to 100 curies now) could be stuffed into a package of constant physical size depending on mission/market. \\
To generate $\sim$10 milliwatts per interstellar probe for decades as described below would require 100 times as much tritium as the battery described below, i.e., 10,000 curies, or 1 gram of tritium.  While no vendors responded directly with price quotations or budget numbers for tritium, in a recent work, it was estimated by one of us that the inflation-adjusted fully-burdened cost of tritium as produced by the (now-decommissioned) Cold-War-era reactor complex at Savannah River would amount to USD $\sim$200,000-300,000 per gram in today's money.\cite{Kennedy2018}  Today, tritium for U.S. nuclear weapons complex is generated by lithium breeder rods at the civilian Tennessee Valley Authority's Watts Bar Nuclear Power Plant.  Tritium for all other purposes, including industrial, is sourced from a fleet of $\sim$20 CANDU reactors, mostly in Canada (hence the name) by periodically ``de-tritiating'' their heavy water (D$_{2}$O) which is used for both a moderator and coolant. The worldwide inventory of tritium is $\le{20}$ kg (which is $\sim$5 orders of magnitude less than the worldwide inventory of $^{90}$Sr below).  The cost of tritium in 2003 was USD $\sim$30,000 per gram; worldwide sales of tritium at that time amount to 100 g. \\
However, there is no room in the extremely-limited interstellar mass budget for radiation shielding or the packaging and pins of traditional chips.  Rather, the $\beta$$^{-}$ emitting layer and the receiving / electricity conversion layers would have to be laid down by direct implantation, and the battery flown without shielding.  Since tritium, an isotope of hydrogen, is normally a gas, the $\beta$$^{-}$-emitting layer of tritium used in the NanoPower$^{TM}$ above would have to be converted into some solid form to be suitable for a long-duration interstellar mission.  While the proprietary details of the Nanopower$^{TM}$ cell pictured below are not publicly available, it should be possible to prepare and work with a mass-efficient solid compound of lithium tritide (LiT), and plate that material directly to a semiconductor such as crystalline silicon.  Tritium's short half-life, 12.3 years, means that power from tritium-fueled cell would decline by about two-thirds on the way to Proxima. \\
Despite its extremely low isotopic mass, tritium's weak $\beta$$^{-}$ ray (0.0186 MeV), makes for a poor power-to-mass ratio, thus providing a milliwatt at best, an order of magnitude less power than the $^{90}$Sr candidate analyzed below. \\
Despite to its many useful industrial applications, tritium has a critical role in thermonuclear weapons, which means its possession and distribution are tightly controlled, albeit not to the severity of fissionable materials ($^{235}$U, $^{233}$U and $^{239}$Pu).  Fortunately, there are viable alternatives subject to less regulation.

\item \underline{Feasible / Near-term Betavoltaics}. 
A method exists now (proven in the laboratory with commercial and medical devices) for directly and continuously \textbf{generating electricity} at current/voltage/power ($\sim$1-10 milliwatts per probe) suitable for deep space missions, by sandwiching commercially-available crystalline silicon (c-Si) $\sim$100-micron-thick photovoltaic layers around a beta emitting layer, with a beta-ray-to-electric conversion efficiency ($\sim$20 percent) almost an order of magnitude better than the thermionic conversion ($\sim$1-5 percent) in RTGs.  This would sufficient to support communication at interstellar range.  Many other photovoltaic materials exist, but have not been put to the test for this application in a laboratory.  (Those experiments ought to commence.)  More exotic much tougher semiconductor photovoltaic materials such as synthetic diamond are worth investigating in this regard. \\ 
Several other isotopes were assessed as source material for betavoltaic power consistent with the extreme mass constraints and mission duration.  We specifically note that these candidates have not been dismissed from consideration for application in interstellar ``chipsats''.  They could be suitable for other missions of different duration to targets either nearer or farther away than Proxima b.  Refining conceptual designs, for example, exactly ``length scale matching'' different types of specially-doped semiconductor to particular thicknesses of each of the three ``runners up'' isotopic fuel candidates ($^{3}$H, $^{32}$Si, $^{63}$Ni) is work that we propose should be taken up in Phase II.\\

In order of isotopic mass, the betavoltaic candidates beyond tritium that we analyzed and compared to tritium are: $^{32}$Si, $^{63}$Ni, $^{90}$Sr.\\
\begin{figure*}[ht]
\centering
\hspace{-0.5cm}
\includegraphics[scale=0.3]{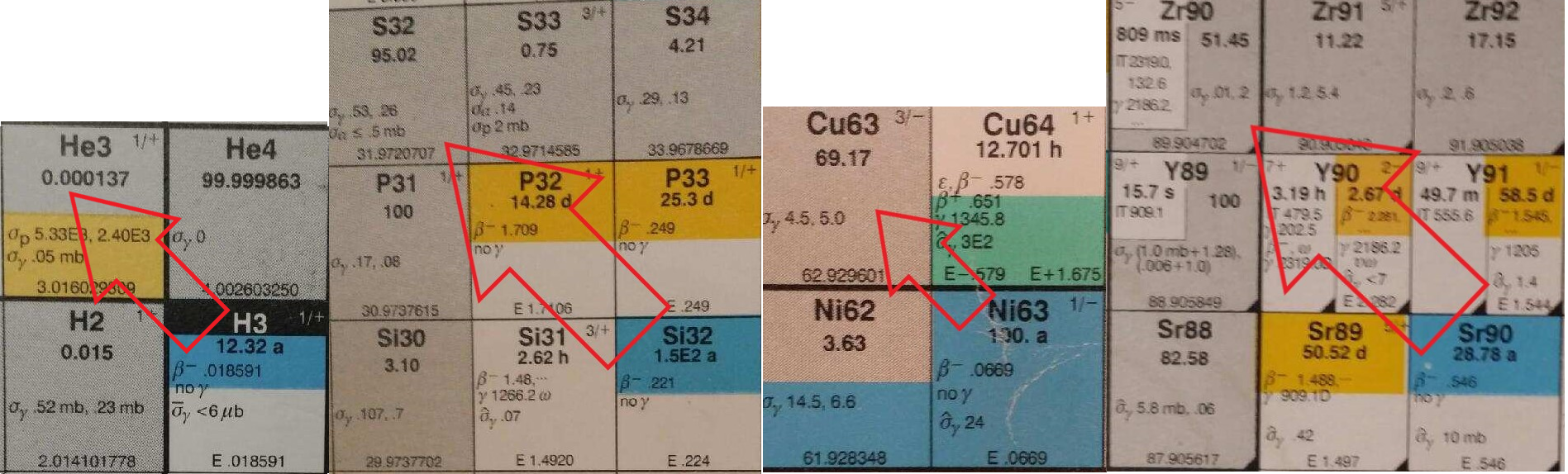}
\caption{Decay chainsr four potential isotopic fuels for interstellar communications, from left to right: $^{3}$H, $^{32}$Si, $^{63}$Ni, $^{90}$Sr.}
\end{figure*}

\textbf{$^{32}$Si} decays via phosphorus-32 to sulfur-32. \\

\begin{equation}
\label{Si32Decay}
^{32}_{14}Si \rightarrow ^{32}_{15}P + \beta ^{-} + \bar{\nu} + (0.221 \mathrm{\ MeV})\ E_{\beta , P}        
\end{equation}

\begin{equation}
\label{P32Decay}
^{32}_{15}P \rightarrow ^{32}_{16}S + \beta ^{-} + \bar{\nu} + (1.709 \mathrm{\ MeV})\ E_{\beta , S}        
\end{equation}

Of all the isotopic fuels we studied, $\sim$2500 distinct radionuclides of the Chart of the Nuclides by Knolls Atomic Power Laboratory, one in particular stands far above the rest: strontium-90.  Unlike tritium or plutonium, strontium-90 is considered a nuisance, a waste product.  There is no shortage of supply, as strontium-90 is one of the most frequent daughter products of uranium-235 fission.  About 6 percent of fission events produce strontium-90, therefore there are approximately 1,000 tons of the stuff sitting in the spent fuel pools of the world's nuclear fleet.  This conjecture was proven by an unofficial quote from the National Isotope Development center (NIDC) at www.isotopes.gov, in which the unit price of strontium-90 came to a negligible USD 0.15 (15 US cents) per full curie (plus a fixed handling fee of USD 57,000).\cite{NIDCemail16Dec22}. This is ten orders of magnitude less than the unit price for silicon-32, and five orders less than the unit price for nickel-63.  Strontium-90 has a 28.8-year half-life, which makes it an excellent match for a 20-year mission to our closest neighbor.  \\
Like $^{32}$Si above, it is at the top of a two-decay chain; in this case, via yttrium-90 to zirconium-90.  It has a powerful combined $\beta$$^{-}$ ray (0.546 + 2.28 MeV), so, despite its moderate isotopic mass, it has by far the best power-to-mass ratio.  Sandwiching semiconductor layers of sufficient thickness (``length scale matching'') around the beta-emitting material converts the primary beta-rays directly into charge, as they interact with p- and n-layers.  Strontium-90 could be prepared as a thin pure metal foil, or as strontium silicate that could be painted onto and should readily adhere to the silicon semiconductor layer.
\begin{figure*}
\centering
\includegraphics[scale=0.5]{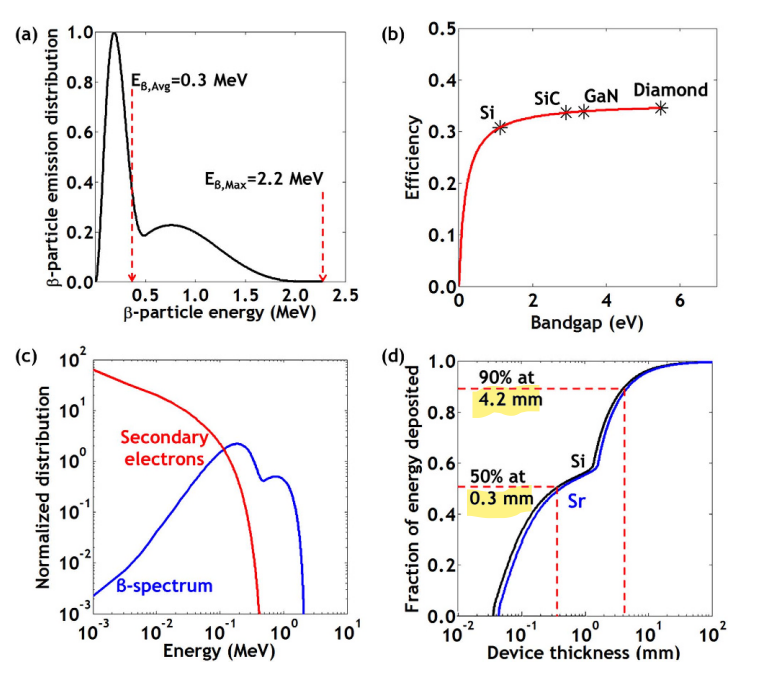}
\caption{Figure from \cite{DixonEtAl2016}.  Length scale matching of a $^{90}$Sr+crystalline silicon betavoltaic device. (a) Normalized energy spectrum of $\beta$ rays emitted by two-step serial decay of $^{90}$Sr then $^{90}$Y. (b) Theoretical efficiency of energy collection in semi-infinite materials. (c) Normalized energy spectrum of secondary electrons generated by $\beta$-spectrum of $^{90}$Sr+$^{90}$Y in comparison to normalized $\beta$-spectrum itself. (d) $\beta$-particle energy deposition in Si and Sr. The ``kink'' near 1.5 mm is due to the higher energy $\beta$-particles (2.2 MeV) from the $^{90}$Y decay. 50 percent of the beta-particle energy is intercepted by 300 microns of material, comparable to thickness of commercial photovoltaic cells.}
\end{figure*}
This has been demonstrated in the lab using a medical electron gun and commercial-grade crystalline silicon.\cite{DixonEtAl2016}.
\begin{figure*}
\centering
\includegraphics[scale=0.42]{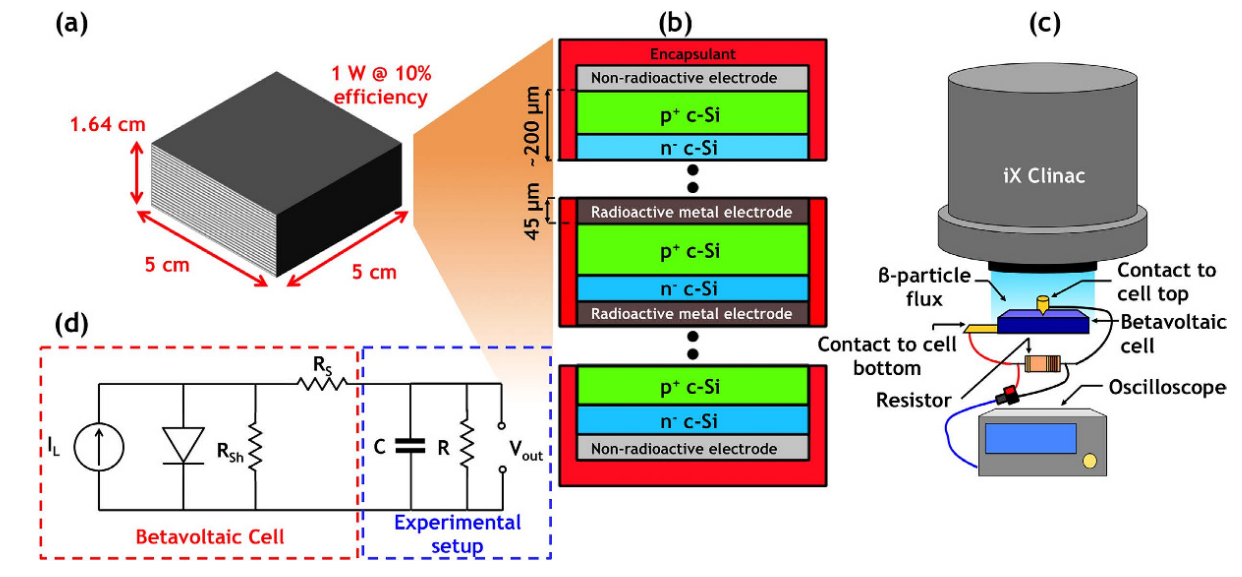}
\caption{Figure from \cite{DixonEtAl2016}.  Betavoltaic device and setup of experiment at Georgia Tech in 2016. (a) Overall dimensions of 1-W device proposed by Dixon \textit{et al.}\cite{DixonEtAl2016}. (b) Cross section. (c) Experimental setup and apparatus to characterize efficacy of betavoltaic cell. (d) Equivalent circuit of setup in (c).  }
\end{figure*}  \\

A betavoltaic cell fueled on this isotope has a specific mass-to-power density of $\sim$32 milligrams per milliwatt, including the ultracapacitor storage (whose mass is but a small fraction of the energy system).  If 30 percent of the total 1-gram mass budget were allocated to energy storage at nuclear densities, then this source could provide about 10 milliwatts of continuous power on the day it is made, and 6 milliwatts of power when flying by Proxima Centauri.  We note that triple-junction silicon cells have achieved photoelectric conversion efficiencies $\ge$40 percent in the laboratory, which would more than double the power output.  Betavoltaic cells fueled by strontium-90 could be available within a decade with a modest technological development effort, of order $\sim$10$^{7}$ dollars. \\
\begin{figure*}
\centering
\includegraphics[scale=0.5]{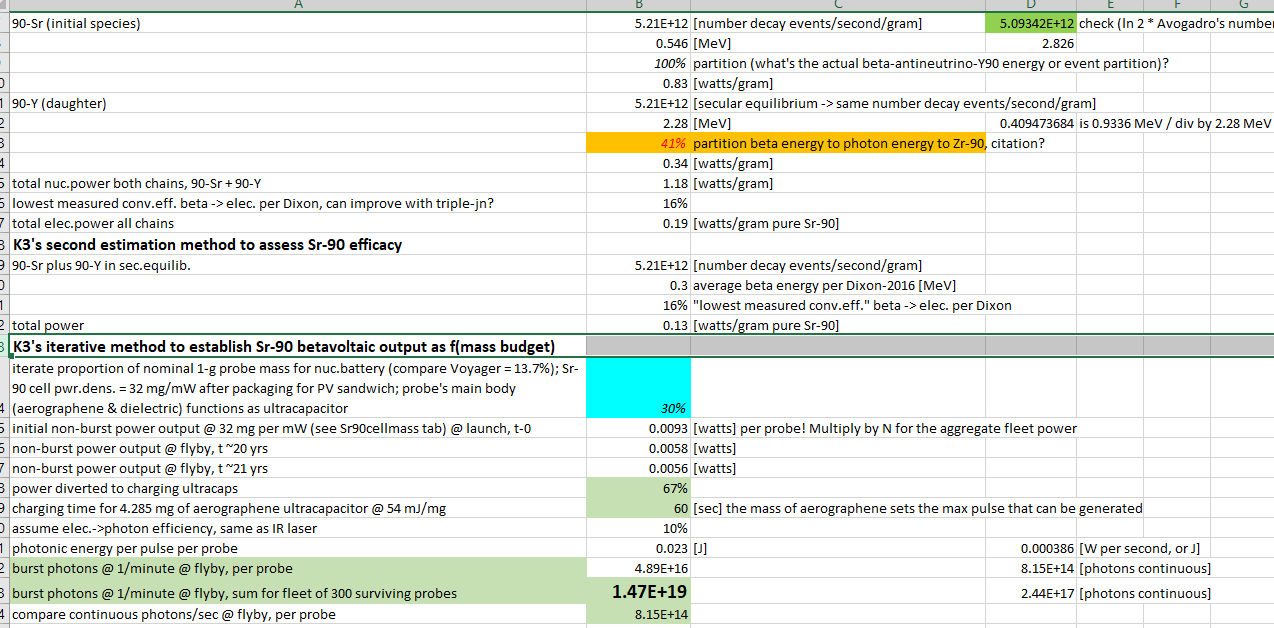}
\caption{Worksheet for Establishing Power Budget by Strontium-90 Betavoltaic Cell.}
\end{figure*}

Seven other $\beta$$^{-}$ emitters and one positron ($\beta$$^{+}$) emitter were found amongst the $\sim$2400 nuclides on the ``Chart of the Nuclides'' published by Knolls Atomic Power Laboratory (\cite{KAPL1996}).  We also note that while double-$\beta$$^{-}$ decay does exist, it is such a rare mode of emission that it occurs on a timescale many orders of magnitude longer than the age of the universe.\\

    \textbf{$^{26}$Al}: the sole positron emitter examined, has two isomers, one with a half-life of 6.3 seconds, much too short for the BTS mission, emits a high-energy positron ($\beta$$^{+}$, 3.21 MeV); the other emits a high-energy $\beta$$^{+}$ at 1.17 MeV, but has a half-life of 71,000 years, far too long for a good power-to-mass ratio.\\
    \textbf{$^{39}$Ar}: has a strong $\beta$$^{-}$ ray (0.565 MeV) and light isotopic mass hence a good power-to-mass ratio, but a light noble gas is difficult to stabilize in solid form without excessive mass penalty; the half-life of 265 years is a much too long for the BTS mission.  \\
    \textbf{$^{85}$Kr}: has two isomers, one with a half-life of 448 hours, far too short, which emits a high-energy $\beta$$^{-}$ (0.840 MeV); the other one also emits a high-energy $\beta$$^{-}$ (0.687 MeV) and has a half-life of 10.76 years, which is a good match for the mission.  The isotopic mass-to-power ratio is acceptable, but a medium noble gas difficult to stabilize in solid form without excessive mass penalty. \\
    \textbf{$^{126}$Sn}: i.e., tin-126, has a modest $\beta$$^{-}$ energy (0.25 MeV) and relatively high isotopic mass, thus a poor power-to-mass ratio.  Furthermore, it comes with a moderately strong $\gamma$ ray (0.085 MeV) that makes it difficult to work with, and its half-life of 250,000 years puts it far out of consideration for the BTS mission. \\
    \textbf{$^{129}$I}: has a modest $\beta$$^{-}$ energy (0.15 MeV), relatively high isotopic mass, thus a poor power-to-mass ratio.  Furthermore, its half-life of 15.7 million years puts it far out of consideration for the BTS mission.  \\
    \textbf{$^{137}$Cs}: like strontium-90 is one of the most significant daughter products of uranium-235 fission, and has a similar half-life, 30.07 years, which makes it an excellent match for the BTS mission.  It has a moderately strong $\beta$$^{-}$ energy (0.514 MeV), but a relatively high isotopic mass, thus a poor power-to-mass ratio; and a hard $\gamma$ ray (0.662 MeV) that would make it difficult to work with. \\
    \textbf{$^{166}$Ho} i.e., holmium-166 has two isomers, one with a half-life of 1.12 days hours, far too short, which emits a high-energy $\beta$ (1.855 MeV); the other one also emits a weak $\beta$ ray (0.065 MeV) and has a half-life of 1200 years when decaying to $^{166}$Er, which is far too long for the mission.\\  
    \textbf{$^{210}$Pb}, i.e., lead-210 has two low-energy $\beta$$^{-}$ emissions: one 0.017 MeV, the other 0.061 MeV.  It has a 22.6-year half-life that makes it an excellent match for the BTS mission, but a high isotopic mass which, combined with the weak betas, gives it a poor power-to-mass ratio.   
\end{enumerate} 
\subsubsection{Harvesting the ISM and Other Exotic Long-Term Options}
\begin{enumerate}
    \item Alphavoltaic cells are less preferred than betavoltaic, due to the orders-of-magnitude greater material damage caused by $\alpha$ particles, and a lesser state of readiness and market availability.  Furthermore, the emission of alpha particles is one of two phenomena that causes something to trigger extreme regulatory precautions. (The other phenomenon is being fissionable, i.e.,``special nuclear material'' (SNM).)  But a probe traveling at relativistic speeds through the ISM must already be designed to withstand a very high radiation dose over its lifetime due to impacts by 20-MeV hydrogen and 80-MeV helium atoms.  Furthermore, the $^{90}$strontium betavoltaic cells have already been physically simulated in the laboratory using a medical 6-MeV electron gun (``Clinac'') fired at commercial solar cells made of crystalline silicon.\cite{DixonEtAl2016}  This is only a factor of 3 less than the 20-MeV hydrogen particles that is the principal constituent of radiation flux induced by traveling through the ISM at 0.2c.  This particular experimental apparatus suggests that may even be possible to utilize the ISM itself as the electrical power source by making the protective leading edge out of a tough semiconductor, thus effectively behaving as a particle-voltaic cell, harvesting the initial investment of launch energy during the entire flight.  
    \item Another very long term possibility for stably storing energy at the ultimate limit of mass-energy could be antimatter inside fullerenes, colloquially known as ``buckyballs'' (the smallest of which is composed of 60 carbons in a spherical lattice, or ``C60'').  Suppose 10 milliwatts of power is needed for 20 years, which is $\sim$6 $\times$ 10$^{8}$ seconds, hence $\sim$6 x 10$^{6}$ joules or 7 x 10$^{-11}$ kg of annihilation, which would take in principle 35 nanograms of antimatter, amounting to order 2 x 10$^{16}$ antiprotons.  One method of storing single antiprotons (not atoms) one at a time in a fullerene trap has been proposed.\cite{Rejcek2003}.  If this is not impossible, then that would require 120 protons and neutrons of $^{12}$C for every antiproton, or $\sim$12 micrograms of C60.  This simple thought experiment does suggest it might be possible to power the BTS ``Starchips'' with very little actual mass.
\end{enumerate}

\section{Precursor Missions}
Earlier work \cite{Parkin-2018-a} produced a point design for a 0.2 c mission carrying 1 g of payload. Recent work by the same author widens the design space to missions having 0.1 mg to 100 kt payload and 0.0001-0.99 c cruise velocity \cite{Parkin-2022-a}.

The first six innovations in our approach could be proved with planned precursor missions, at velocities up to 0.01c.  The team has done and continues to do extensive study of such 
missions  \cite{Hein-et-al-2019-a,Hibberd-et-al-2019-a,Hein-et-al-2020-b,Hibberd-Hein-2020-a,Eubanks-et-al-2020-b}.  While precursor missions could go to any body in the solar system, we further propose a targeted flyby of the ISO 1I/'Oumuamua.  Much of the scientific discussion of this enigmatic object concerns its anomalous acceleration, which is associated with its currently very large ephemeris error.  The simple detection of 1I at the one-pixel level in a Starshot precursor chipsat would provide a substantial scientific return and thus enable future missions.  A second precursor mission could be tightly targeted and could realistically return images of 1I to Earth by 2035-2040.  Other fast-flyby precursor missions to celestial bodies such as trans-Neptunian objects (TNOs) beyond the Edgeworth-Kuiper Belt in the outer solar system could return valuable science far more rapidly than conventional deep-space missions presently taking near lifetimes.\\
The existence of Interstellar Objects (ISOs) visiting the Solar System has been predicted for many years (e.g. \cite{Sekanina-1976-a,Stern-1990-a}). We live in an interesting epoch were two of them have been found (1I/'Oumuamua) and (2I/Borisov) and signs of other visitors have been proposed \citep{Siraj-Loeb-2018-a,Froncisz-et-al-2020-a}. In addition, there are prospects to detect one or several of them per year with the Vera Rubin Observatory (or LSST) starting in 2022 \citep{Trilling-et-al-2017-a}. Ground-based telescopes will not be able to give reliable answers to questions about the origin, chemical composition, mineral structure, size, shape of these interstellar visitors - that will require \textit{in situ} exploration.

\section{Conclusions}
\begin{enumerate}
\item Interstellar communications are achievable with gram-scale spacecraft using the spacecraft swarm techniques introduced here if an adequate energy source, clocks and a suitable communications protocol exist.  Our proposed communications system will filter out Proxima’s background photon flux using a combination of frequency and time bandpasses, and also by using a short wavelength at the source (432 nm, is red-shifted to 539 nm at Earth) where the Proxima flux is very weak.
\item A swarm of 100s to 1000s of spacecraft of order $\sim$100,000 km diameter can be established en route to Proxima b!  This can be accomplished with a combination of a gross ``time-on-target'' (ToT) technique, consisting of modulating the initial velocity of each probe by the launch laser such that the tail catches up with the head, relative to each other, and a finer ``velocity-on-target'' (VoT) technique based on controlled drag imparted by the interstellar medium (ISM) by altering the attitude of individual probes with respect to the ISM, thus keeping swarm together in relative and absolute position once formed.  This will take time, but time is something we have a lot of.
\item Attitude adjustment is also necessary to minimize the severe erosion ($\sim$100 microns per day) by particle impacts and extremely high radiation dose (gigarads) induced by traveling through the ISM at 0.2c.  To minimize frontal area, hence dose and erosion, we anticipate flying edge-on most of the way, without rotation, hence distinct leading and trailing edges, even though the probe is mostly symmetric.
\item An operationally-coherent (if not phase-coherent) i.e. synchronized swarm supported by forecast-state-of-the-art space-rated clock metrology onboard (10$^{-13}$) and supported by our forecast-by-2050 state-of-the-art clock metrology at Earth (10$^{-19}$ or better), and existing time- and frequency-bandpass filtering, can feasibly support the transmission of sufficient signal photons for reception at Earth of $\gtrsim$ 100K bytes of science data.  Space-rated, chip-scale atomic clocks (CSAC) are already commercially available at very low cost (\$2K) today, although not yet in the miniscule mass that would fit it a probe's budget.
\item At least one candidate construction metamaterial, aerographene, exists today, that has sufficiently low density to satisfy the payload mass constraint, yet sufficient mechanical strength and other qualities to serve as the basis for the main body of a probe body has a variable density, tailored for the particular mechanical requirements. 
\item A true optical phased array of elements (247) can fill the full diameter (4 m) of an individual probe, which would support both imaging of the target and return of the science data to Earth at far higher signal-to-noise ratio than currently thought.
\item Intra-swarm (probe-to-probe) communication can be established by an independent orthogonal system of lightweight efficient infrared (12,000 nm) laser-beacons of modest aperture (20 mm). 
\item At least one near-term candidate method of \textbf{storing primary energy} onboard in sufficiently compact form, i.e., nuclear energy density.  \\
In addition, a method exists now for organically and continuously \textbf{generating electricity} therefrom, at current/voltage/power ($\sim$3-10 milliwatts per probe) sufficient for interstellar coms.  Betavoltaic batteries fueled by the isotope strontium-90 ($^{90}$Sr), sandwiched inside commercially-available crystalline silicon (c-Si) photovoltaic material, could be available in a decade (by $\sim$2032) with modest technology and manufacturing development of order $\sim$10$^{7}$ dollars.  $^{90}$Sr is a waste product of uranium fission. It has the additional logistical advantages of not being regulated as a ``special nuclear material'', and its supply is practically unlimited ($\sim$1,000 tonnes), seven orders of magnitude greater than is needed to achieve the Breakthrough Starshot mission.  Already, a betavoltaic cell fueled with tritium ($^{3}$H) is commercially available today, albeit at an energy density or power output four orders of magnitude less than would be necessary for a 1-gram interstellar probe.  \\
A method exists now of \textbf{storing electricity} as high mass efficiency, well as discharging it in the necessary time (nanoseconds), namely ultracapacitors.\\
The entire power system falls within the available mass budget, $\sim$330 milligrams out of 1 gram payload limit, or 30 percent partition, which is of the same order as the 15 percent allocation in the Voyager probes of two generations ago.  \\
\item The probes we propose cannot be manufactured by conventional methods based on subtraction and assembly of subassemblies, but rather a monolithic process involving a combination of additive manufacturing and wafer-scale integration which however must be scaled up by order of magnitude in extent, from 30 cm at present to 4 m.
\item  The $\sim$100-GW drive laser can feasibly be repurposed for a number of important suporting tasks after its principal use for launching the fleet:\\
    serving as a beacon for individual probes to point at when communication back toward Earth\\
    for direct communication from Earth to the fleet for the entire journey in order to provide astronomical and astrometric updates about the target, and synchronization for position-navigation-timing (PNT), albeit at an increasing temporal latency (4.3 years at Proxima b)\\
    to continuously calibrate and optimize the main communication channel between the fleet and the Earth over the course of the entire cruise phase, and after during the data return phase.\\
    as an ``interstellar flashlight'' finding small bodies at encounter, as well as provide an additional marginal but possibly useful controlled source of illumination for photography of dark objects.\\
    as a adjunct piece of scientific equipment namely a monochromatic light source for spectroscopy by the fleet during flyby.
\item \textit{Sine qua non}: In order for any of this to happen, we have identified a fundamental operational unknown that must be solved well in advance: accurately determining the orbital position of Proxima b at least 8.6 Earth years ($\sim$300 revolutions) ahead of launch.  We find that this can be overcome at reasonable cost through the targeted use of gravitational microlensing observed by small telescopes in low Earth orbit (LEO).  
\end{enumerate}

\section{Acknowledgments}

The Breakthrough Starshot Foundation funded this work, which was also supported and executed by the Institute for Interstellar Studies (US), the Initiative for Interstellar Studies (UK), and Space Initiatives Inc.

\bibliographystyle{cas-model2-names}

\bibliography{Breakthrough}



\end{document}